\documentclass[fleqn,usenatbib]{mnras}

\usepackage{newtxtext,newtxmath}
\usepackage{tcolorbox}

\usepackage[T1]{fontenc}
\usepackage{enumitem}
\usepackage{soul}

\DeclareRobustCommand{\VAN}[3]{#2}
\let\VANthebibliography\thebibliography
\def\thebibliography{\DeclareRobustCommand{\VAN}[3]{##3}\VANthebibliography}

\usepackage{graphicx}	
\usepackage{amsmath}	
\usepackage{float}
\usepackage{stfloats}

\title[DEVILS: Morphological vs close-pair merger fractions]{Deep Extragalactic VIsible Legacy Survey (DEVILS): Morphologically-selected galaxy merger fractions and their direct comparison to close-pair samples}

\author[Fuentealba-Fuentes et al.]{Melissa F. Fuentealba-Fuentes$^{1}$\thanks{E-mail: melissa.fuentealbafuentes@research.uwa.edu.au},
Luke J. M. Davies$^{1}$,
Aaron S. G. Robotham$^{1}$,
Sabine Bellstedt$^{1}$, 
\newauthor
Claudia D. P. Lagos$^{1}$,
Michael J. I. Brown$^{2}$,
Malgorzata Siudek$^{3,4}$
\\
$^{1}$ICRAR, The University of Western Australia, 35 Stirling Highway, Crawley, WA 6009, Australia
\\
$^{2}$School of Physics, Monash University, Clayton, VIC 3800, Australia
\\
$^{3}$Instituto Astrofisica de Canarias, Av. Via Lactea s/n, E38205 La Laguna, Spain
\\ 
$^{4}$Institute of Space Sciences (ICE, CSIC), Campus UAB, Carrer de Can Magrans, s/n, 08193 Barcelona, Spain}

\date{Accepted XXX. Received YYY; in original form ZZZ}

\pubyear{2025}

\begin{document}
\label{firstpage}
\pagerange{\pageref{firstpage}--\pageref{lastpage}}
\maketitle

\begin{abstract}
Galaxy mergers are a central driver of galaxy evolution across cosmic time, and thus, quantifying their frequency is critical for constraining hierarchical models of galaxy formation. Motivated by the need to robustly quantify these fractions and their evolution, we build on our previous close-pair analysis by exploring morphological identification techniques within the Deep Extragalactic VIsible Legacy Survey (DEVILS), using the D10 (COSMOS) field, which covers an area of $1.47$ deg$^2$. While close-pairs trace the early stages of galaxy interactions, morphological methods probe more advanced phases of the merging process, including systems with disturbed structures and post-merger remnants. We present galaxy merger fractions over the redshift range $0.2 < z < 0.9$ using visual classification and automated identification based on non-parametric statistics: concentration ($C$), asymmetry ($A$), smoothness ($S$), Gini ($G$), and $M_{20}$, applied to HST/ACS imaging. To enhance the detection of subtle structural perturbations, we measure asymmetry on unsharp-masked images. 
We find relatively little overlap between visually and automatically identified samples, which highlights their distinct sensitivities and limitations. Moreover, galaxy merger fractions derived from morphological disturbances are consistently higher than those from close-pair counts at all redshifts. This potentially reflects how each method probes different stages of the merger process, with distinct observability timescales, as well as the fact that morphologically disturbed galaxies, at a given redshift, are typically the later-stage descendants of close-pairs from earlier epochs. This comparison allows us to examine systematic differences between identification techniques and assess how they impact the observed evolution of the galaxy merger fraction.

\end{abstract}

\begin{keywords}
galaxies: evolution – galaxies: interactions
\end{keywords}



\section{Introduction} \label{Introduction}

Galaxy mergers are a natural outcome of a hierarchical, cold dark matter (CDM) universe. They play a fundamental role in galaxy evolution, acting as a primary mechanism for mass redistribution and the reshaping of galaxy populations. Numerous studies have explored the impact of mergers on triggering star formation \citep[e.g.][]{Hernquist_1989, Ellison_2008, Davies_2015}, fueling AGN activity \citep[e.g.][]{Di_Matteo_2005, Hopkins_2008, Dekel_Burkert_2014, Ellison_2019, Erostegui_2025}, and driving morphological transformation \citep[e.g.][]{Lotz_2008b, Bournaud_2011, Lagos_2017, Martin_2018, Lagos_2018, Lagos_2022}. Equally important is quantifying the frequency of these events, since merger fractions and rates offer key observational constraints for hierarchical galaxy formation models and their ability to reproduce the observed evolution of galaxies across cosmic time. Galaxy merger fractions quantify the proportion of galaxies within a given sample that show evidence of merging at the time of observation. These quantities are generally easier to estimate than merger rates, which describe the frequency of merger events over time, typically measured in mergers per gigayear. Merger fractions can be converted into merger rates by dividing by a merger observability timescale, which represents the period during which a merger can be identified before the galaxies fully coalesce. These timescales are usually derived from simulations and represent a significant source of uncertainty.

Measuring these fractions and rates requires the robust identification of merging systems. However, this remains a complex task, and a variety of techniques have been developed, including visual classification \citep[e.g.][]{Kartaltepe_2015}, quantitative morphological indicators \citep[e.g.][]{Abraham_2003, Conselice_2003, Lotz_2004}, and close-pair selection \citep[e.g.][]{Patton_2000, Kartaltepe_2007, Robotham_2014, Duncan_2019, Fuentealba-Fuentes_2025}. Each method possesses its own strengths and inherent limitations, based on specific observational requirements and the stages of the merging process to which they are sensitive, from nearly unperturbed pairs to post-merger remnants.

The close-pair technique identifies galaxies in the early stages of galaxy interactions, where systems are dynamically close and likely to merge in the future, yet remain spatially distinct. Close-pairs are defined as galaxies that are close both in projected spatial separation ($r_\mathrm{sep}$) and in radial velocity ($\Delta{v}$). Typically, studies adopt thresholds of $r_\mathrm{sep} < 20-50$ kpc and $\Delta{v} < 500$ km s$^{-1}$, for spectroscopic samples \citep[e.g.][]{Patton_2002, Robotham_2014}. Ideally, constraining merger fractions across all redshifts would require large, highly complete spectroscopic samples. However, such completeness has so far only been achieved at low redshift, as demonstrated by \citet{Robotham_2014} using GAMA \citep[][]{Driver_2011} for $0.05 < z < 0.2$. At higher redshifts, studies often rely on photometric data alone. While photometry provides the necessary depth and volume, the lack of precise radial velocity information requires adopting a much wider velocity selection window. Without properly accounting for these larger uncertainties, photometric samples are susceptible to contamination from chance projections, which can significantly bias the estimated merger fractions \citep[e.g.][]{Patton_2000}. Nevertheless, these projection effects can be modeled and statistically corrected by applying control samples or by using mock catalogs derived from cosmological simulations to quantify the frequency of line-of-sight alignments. 
 
To mitigate these observational biases, spectroscopic surveys providing precise redshifts and high completeness are essential, particularly at higher redshifts. In this context, our previous work \citet{Fuentealba-Fuentes_2025} addressed these challenges by measuring galaxy merger fractions and rates up to $z \sim 0.9$, using the Deep Extragalactic VIsible Legacy Survey (DEVILS; \citealt{Davies_2018, Davies_2021}). We first presented a robust measurement of the major close-pair fraction and rate using a well-defined spectroscopic sample over 0.2 $< z < 0.34$ and a stellar mass range of log$_{10}$($M_\star$/$M_\odot$) = 10.66 $\pm$ 0.25 dex, ensuring stellar mass and colour completeness across the entire redshift range. We then extended this analysis to 0.2 $< z <$ 0.9 by combining the full redshift probability distributions of galaxies with high-quality spectroscopic, photometric, or grism measurements, within the same stellar mass bin. 

Building on this work, we now focus on an alternative approach for identifying mergers: morphological techniques and directly compare to our close-pairs results. These methods include visual classification, which identifies mergers through signatures of structural perturbations such as common envelopes, tidal tails, bridges, or double nuclei, as well as automated diagnostics, which quantitatively distinguish isolated galaxies from interacting systems based on their light distributions. Commonly used non-parametric statistics include concentration ($C$), asymmetry ($A$), and smoothness ($S$) \citep[also known as $CAS$ parameters,][]{Bershady_2000, Conselice_2003}, as well as the Gini ($G$) coefficient and the second-order moment of the brightest 20$\%$ of galaxy light, $M_{20}$ \citep[][]{Abraham_2003, Lotz_2004}. Typically, these diagnostics are combined to identify a wider range of merger stages and to reduce the biases of the individual parameter spaces. Visual classification captures complex tidal features, while automated diagnostics provide a reproducible, statistical approach that reduces human subjectivity.

A key challenge in reconciling these methods with close-pair studies lies in their different observability timescales ($T_\mathrm{obs}$). Simulations by \citet{Lotz_2008b} show that close-pair selections with small projected separations \citep[$r_\mathrm{sep}<20\,h^{-1}\,\mathrm{kpc}$, as used in][]{Fuentealba-Fuentes_2025} typically probe timescales of $\sim0.1$--$0.6$ Gyr, while morphology-based diagnostics span $\sim0.2$--$1$ Gyr depending on the specific statistic and gas fraction of the interaction. Therefore, the observability timescales associated with the two methods can be comparable or differ significantly depending on the adopted merger selection criteria. 
Moreover, morphology-based diagnostics are typically biased toward major mergers (stellar mass ratios down to $1:3$), which drive the most significant structural transformations and mass redistribution \citep[e.g.][]{Lagos_2018}. In contrast, minor mergers primarily contribute to the growth of galaxy bulges and disk thickening, often producing subtle asymmetries that fall below detection thresholds \citep[e.g.][]{Lotz_2010, Bottrell_2024}. At higher redshifts, morphological identification is further complicated by decreasing spatial resolution and cosmological surface brightness dimming, which scales as $(1 + z)^4$ \citep[][]{Tolman_1930, Tolman_1934}. This dimming effect significantly reduces the visibility of faint tidal features, potentially leading to an underestimate of the galaxy merger fraction at $z > 0.5$.

To enable a direct comparison between close-pair and morphology-based merger fractions, it is therefore essential to use samples drawn from the same survey volume and defined over comparable stellar mass ranges. This is the approach we adopt in this paper.
We present estimates of galaxy merger fractions over the redshift range $0.2 < z < 0.9$, using visual classifications and automated morphological identification techniques applied to high-resolution Hubble Space Telescope (HST) imaging. The galaxies are drawn from the Deep Extragalactic VIsible Legacy Survey \citep[DEVILS;][]{Davies_2018, Davies_2025e}, which provides a highly complete sample of galaxies. To better detect faint tidal features, we use unsharp-masking techniques to isolate small-scale structures from the global light distribution of the galaxy, thereby increasing the sensitivity of our asymmetry measurements to low-surface-brightness disturbances. This work allows a direct comparison with the galaxy merger fraction estimates presented in \citet{Fuentealba-Fuentes_2025}, which were based on close-pair galaxies over the same redshift range and using the same major merger mass-ratio definition. By using the same field and sample, we aim to determine if the evolutionary trend of the galaxy merger fraction is consistent across different identification techniques, even if the absolute values differ due to the distinct observability timescales of each method.

Throughout this paper, we assume a $\Lambda$CDM cosmology with parameters corresponding to total matter and $\Lambda$ densities of $\Omega_m$ = 0.3, and $\Omega_{\Lambda}$ = 0.7, and a Hubble constant at $z=0$ of H$_0$ = 70 kms$^{-1}$ Mpc$^{-1}$.

\section{Non-parametric morphological measurements}

To quantitatively characterize galaxy structure and identify merger candidates, we use non-parametric statistics such as asymmetry, concentration, smoothness, Gini, and $M_{20}$ to capture the irregular and disturbed morphologies typically observed in galaxy interactions. In this section, we describe each of these diagnostics in detail.

\subsection{Asymmetry}

The asymmetry parameter ($A$) measures the difference between the original galaxy image and its counterpart rotated by 180$^{\circ}$ \citep{Abraham_1996, Conselice_2000}. It is calculated as follows:

\begin{equation}
    A = \frac{\sum |I - I_{180}|}{\sum |I|} - A_{\mathrm{bgr}} ,
    \label{eq:A_def}
\end{equation}

\noindent where $I$ is the original image, $I_{180}$ is the image rotated by 180$^{\circ}$ about the galaxy's center, and $A_{\mathrm{bgr}}$ represents the asymmetry of the background, calculated from an empty sky region to account for noise. Asymmetry is typically higher in galaxy mergers than in non-mergers, as tidal forces and gravitational interactions distort the galaxy’s structure, producing irregularities, tidal tails, and asymmetric light distributions.

\subsection{Concentration}

The concentration index ($C$) is usually defined as \citep{Bershady_2000, Conselice_2003}: 

\begin{equation}
    C = 5 \log_{10} \left(\frac{r_{80}}{r_{20}} \right),
\end{equation}

\noindent where $r_{20}$ and $r_{80}$ are the radii of circular apertures containing $20 \%$ and $80 \%$ of the galaxy’s total light, respectively. The total flux is measured within $1.5\,r_\mathrm{petro}$, as used in recent studies \citep{Conselice_2003, Lotz_2004}. The centre of the aperture corresponds to the point that minimizes the asymmetry index. For mergers, concentration is often lower because interactions disrupt the central light peak and spread flux into tidal features and off-centre clumps.

\subsection{Smoothness}
The smoothness ($S$) or also called "clumpiness" index is calculated by subtracting a smoothed version of the galaxy image, obtained using a boxcar filter of width $\sigma$, from the original image \citep[][]{Conselice_2003}, following:
\begin{equation}
    S = \frac{\sum I - I^{S}}{\sum I} - S_{\mathrm{bgr}} ,
\end{equation}
\noindent where $I$ is the original image, $I^{S}$ is the smoothed image, and $S_{\mathrm{bgr}}$ represents the average smoothness of the background. Larger values of $S$ indicates galaxies that are less smooth (i.e. more clumpy). The sum is carried out over all pixels at radii between $\sigma$ and $1.5 r_{\mathrm{petro}}$ from the centre that minimizes the asymmetry. In merger studies, the smoothness is commonly used alongside the asymmetry index to help distinguish between morphologically disturbed mergers and clumpy, non-merging star-forming galaxies.

\subsection{Gini coefficient}

The Gini coefficient ($G$) measures the inequality in the distribution of pixel intensities within a galaxy image. Originally developed in economics to quantify wealth inequality in a population, it was first applied to astronomy by \citet{Abraham_2003}. It is calculated as:

\begin{equation}
    G = \frac{1}{\overline{|X|}n(n-1)} \sum_{i=1}^n (2i - n -1)|X_i|,
\end{equation}

\noindent where $|X_i|$ is the intensity of the $i^\text{th}$ pixel, $\overline{|X|}$ is the mean of the absolute pixel intensities, and $n$ is the total number of pixels in the galaxy as defined by the segmentation map. For mergers, the Gini coefficient is usually high because interactions create an uneven light distribution.

\subsection{$M_{20}$ statistic}

The $M_{20}$ statistic \citep{Lotz_2004} measures the second moment of a galaxy’s brightest regions, containing $20$$\%$ of the total flux, relative to the total second-order central moment, $\mu_\mathrm{tot}$, which is defined as

\begin{equation}
    \mu_{\mathrm{tot}} = \sum_{i=1}^{n} \mu_i \equiv \sum_{i=1}^{n} I_i [(x_i - x_c)^2 + (y_i - y_c)^2 ],
\end{equation}

\noindent where $I_i$ are the pixel flux values, $(xc, yc)$ is the galaxy’s centre, and the sum is carried out over all the pixels identified by the Gini segmentation map. The centre $(xc, yc)$ corresponds to the point that minimizes $\mu_\mathrm{tot}$, which represents the centroid of the pixels labelled in the segmentation map. 

Essentially, $M_{20}$ represents the spatial spread of the brightest pixels across the galaxy. It is obtained by sorting the pixels by flux and summing $\mu_i$ over the brightest pixels until their cumulative flux equals $20$$\%$ of the galaxy’s total flux. The result is then normalized by $\mu_\mathrm{tot}$, so that $M_{20}$ is ultimately defined as:

\begin{equation}
    M_{20} \equiv \log_{10} \left( \frac{\sum_i \mu_i}{\mu_\mathrm{tot}} \right), \mathrm{while}
    \sum_i I_i < 0.2\, I_\mathrm{tot},
\end{equation}

\noindent where $I_\mathrm{tot}$ is the total flux of the pixels identified by the segmentation map. Mergers generally exhibit high (less negative) $M_{20}$ values, indicating that a significant fraction of the galaxy’s light is contained in off-center bright regions, such as double nuclei.

\section{Data} \label{Data}

\subsection{The sample}

The Deep Extragalactic VIsible Legacy Survey \citep[DEVILS;][]{Davies_2018, Davies_2025e} is an optical spectroscopic survey at the Anglo-Australian Telescope (AAT). DEVILS was designed to achieve high spectroscopic completeness to $Y$ $\lesssim 21.2$ mag in the three well-studied extragalactic fields: D10 (COSMOS), D02 (ECDFS) and D03 (XMM-LSS), covering a total area of $\sim$ 4.5 deg$^2$. This highly complete sample at intermediate redshifts allows for the robust characterization of group and pair environments in the distant Universe. 

In this work, we focus on a sample of galaxies in the D10 (COSMOS) field to allow a direct comparison between two methods for identifying and measuring galaxy merger fractions: the morphological classifications presented in this work and close-pair statistics reported in \citet{Fuentealba-Fuentes_2025}, which also used the D10 field. D10 was selected in our previous work because it is the most spectroscopically complete of the DEVILS fields (greater than 85$\%$), covering $\approx$1.47 deg$^2$, and incorporating several existing spectroscopic programs, along with high-robustness, high-accuracy photometric redshifts from both the Physics of the Accelerating Universe Survey \citep[PAUS;][]{Alarcon_2021, Cabayol_2023, Serrano_2023} and the Cosmic Evolution Survey \citep[COSMOS2015;][]{Laigle_2016}. 

The D10 field also overlaps with high-resolution observations from the Hubble Space Telescope (HST) Advanced Camera for Surveys (ACS), providing the imaging quality required for our visual and automated identification of mergers. This field and its HST/ACS imaging have already been used for structural and morphological studies of DEVILS galaxies \citep{Hashemizadeh_2021, Hashemizadeh_2022, Cook_2025}. The HST/ACS mosaic is an image in the F814W filter (AB system), with dimensions $158{,}000 \times 137{,}500$ pixels and a pixel scale of $\sim 0.03''$ per pixel. The limiting point-source depth ($5\sigma$) is $27.2$ AB mag \citep[][]{Koekemoer_2007}. The final mosaic was produced by median combining individual exposures using \texttt{SWarp} \citep[][]{Bertin_2002} and calibrated to an AB magnitude zero-point of $21.1$.

\subsection{Stellar mass/redshift selection}

In \citet{Fuentealba-Fuentes_2025}, we used two galaxy samples: one consisting of galaxies with spectroscopic redshifts only (the spec-$z$ sample) from several existing spectroscopic programs \citep[see DEVILS first data release described in][for further details]{Davies_2025e}, and a second including galaxies with spectroscopic, photometric, and grism redshifts (the photo$+$spec-$z$ sample, for simplicity). The photometric redshifts were drawn from the Physics of the Accelerating Universe Survey \citep[PAUS;][]{Alarcon_2021, Cabayol_2023, Serrano_2023} and the Cosmic Evolution Survey \citep[COSMOS2015;][]{Laigle_2016}. The grism redshifts were obtained from 3D-HST \citep[][]{Brammer_2012, Skelton_2014, Momcheva_2016} and the PRism MUlti-object Survey \citep[PRIMUS;][]{Alison_2011, Cool_2013}. 

The key difference between the two samples is that we applied a completeness limit to the spec-$z$ sample, restricting the study to the redshift range $0.2 < z < 0.34$. Within this sample, we measured the major close-pair fraction in a stellar-mass bin of $\mathcal{M}^{*} \pm 0.25$ dex, where $\mathcal{M}^{*}$ ($= 10^{10.66} M_\odot$) is the characteristic mass marking the knee of the galaxy stellar mass function (GSMF) and corresponds to the peak in the number density of close-pairs at $z = 0$ \citep[][]{Patton_2008, Robotham_2014}. In particular, \citet{Thorne_2021} show that $\mathcal{M}^{*}$ does not evolve with redshift out to $z \sim 1$. 

Within this stellar mass/redshift bin, we measured the major close-pair fractions while ensuring completeness in both stellar mass and colour. This constraint is crucial because DEVILS is an observed-frame, magnitude-limited (Y-band) survey, such that at higher redshifts the galaxy sample becomes increasingly biased toward intrinsically brighter and more massive systems. Such effects introduce selection biases and incompleteness that can affect the robustness of the inferred galaxy merger rate. For further details of this selection, see \citet{Fuentealba-Fuentes_2025}. For the photo$+$spec-$z$ sample, we estimated the close-pair fractions within the same stellar mass bin defined for the spec-$z$ sample, but without applying any completeness limit. This allowed us to extend the analysis from $z \sim 0.2$ up to $z \sim 0.9$.

\begin{figure}
    \centering
	\includegraphics[scale=0.53]{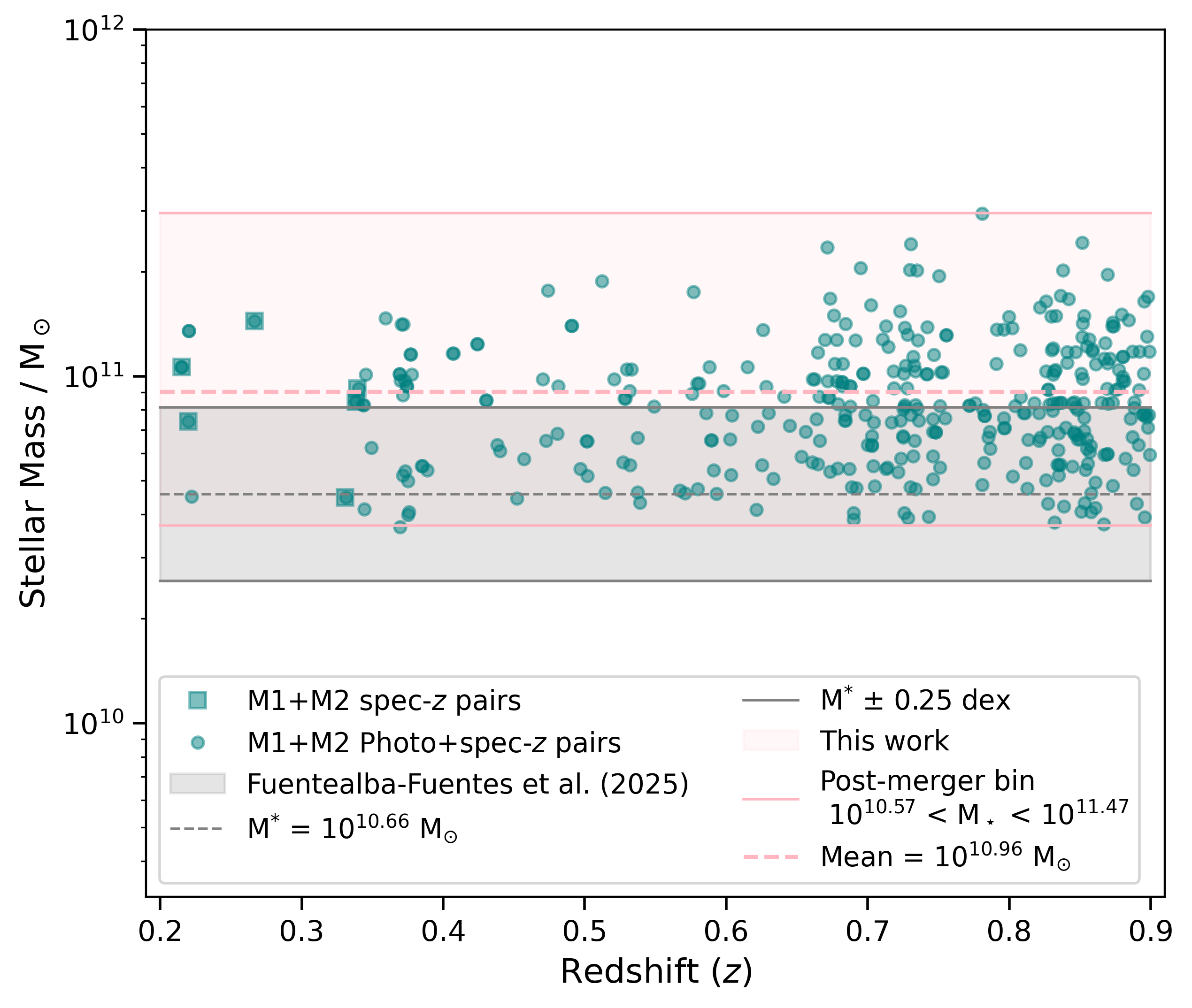}
    \caption{The stellar mass/redshift bin used to estimate the galaxy merger fractions in the D10 (COSMOS) field. The grey region shows the bin adopted in \citet{Fuentealba-Fuentes_2025}, which ensures stellar mass and colour completeness. The pink region indicates the bin used in this work, defined to enclose the resulting stellar masses (green squares and circles), assuming that all the close-pairs identified in \citet{Fuentealba-Fuentes_2025} have merged. This choice allows a direct comparison between galaxy merger fractions inferred from close-pairs and those obtained through morphological identification, by selecting the expected stellar masses of merger remnants.}
    \label{fig:BinComparison}
\end{figure}

In this work, we define a single sample that includes galaxies with spectroscopic, photometric, and grism redshifts. The sample covers a redshift range of $0.2 < z < 0.9$, and the stellar mass bin is based on, but not identical to, the stellar mass range used in \citet{Fuentealba-Fuentes_2025}. To allow a meaningful comparison between this morphology-based study and the previous close-pair analysis, we assume that the systems identified here as ongoing or post-merger interactions would previously have been observed as close-pairs in our first analysis. Therefore, we define the stellar mass bin to correspond to the expected masses of the remnant galaxies, i.e. the merger products of the close-pairs analysed in \citet{Fuentealba-Fuentes_2025}. This results in a stellar mass range of $10^{10.57} M_\odot < M_\star < 10^{11.47} M_\odot$ (see Figure \ref{fig:BinComparison}). The galaxies in this stellar mass bin may thus represent a later stage of similar systems to those in \citet{Fuentealba-Fuentes_2025}, and the corresponding galaxy merger fraction may reflect this temporal evolution. We also tested an alternative stellar mass bin centered on the mean of our new range ($M_\star = 10^{10.96} M_\odot$) but using the same width from the close-pair analysis ($\pm 0.25$ dex) and found no significant change in the visually or automatically identified galaxy merger fractions, which remained consistent across all redshifts.

\begin{figure}
    \centering
    \includegraphics[scale=0.137] 
    {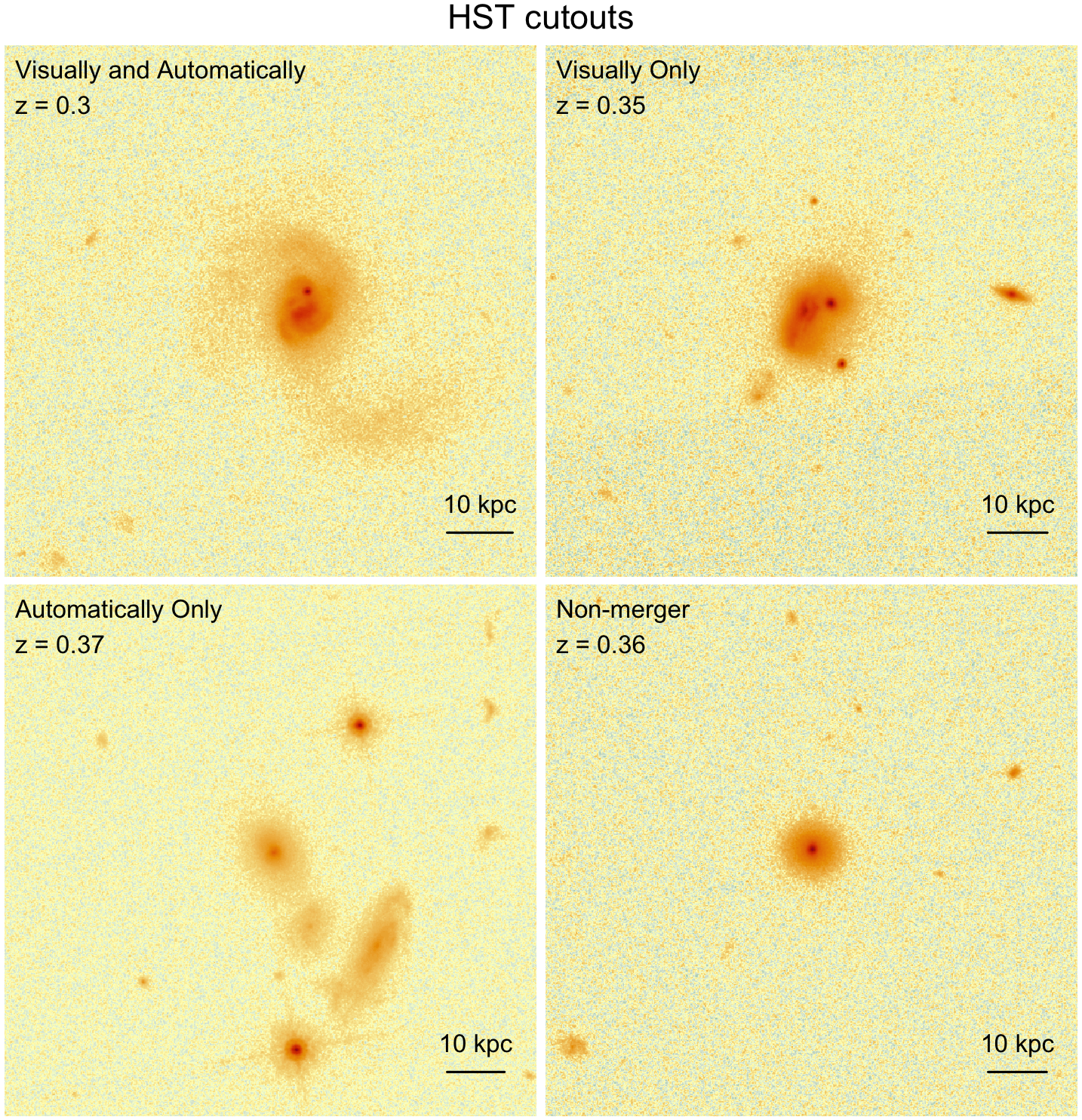}
    \caption{Examples of cutouts from the HST/ACS mosaic. Panels are categorized by merger classification: both visual and automated, visual only, automated only, and non-mergers. Here, "Visual" indicates galaxies identified as mergers by all four classifiers, while no visual confirmation indicates galaxies classified as non-mergers by all.}
    \label{fig:cutout}
\end{figure}

The final sample contains 7,006 galaxies, after excluding 447 systems for which no imaging is available because they lie outside the area covered by the HST/ACS mosaic. For each galaxy, we create a cutout image from the HST/ACS mosaic with dimensions of $600 \times 600$ pixels, corresponding to $18'' \times 18''$ on the sky (see Figure \ref{fig:cutout} for examples). These cutouts were used for both the visual and automated classification of major mergers.

\section{Visual classification} \label{Visual}

We first identified mergers by visually selecting galaxies that show evidence of an ongoing or recent major interaction (stellar mass ratio $\geq 1:3$), specifically targeting disturbed systems with features such as common envelopes, tidal tails, bridges, or multiple nuclei. Potential close-pairs without visible perturbations in their light distribution were not considered major mergers, maintaining a clear distinction from the close-pair selection in \citet{Fuentealba-Fuentes_2025}. This visual classification was carried out independently by four of the authors of this work, using the HST/ACS cutouts of $600 \times 600$ pixels, without providing information on the redshifts of the sources. We first discussed a small subset of galaxies to calibrate our classification criteria, specifically deciding which features should be considered mergers and which apparent close-pairs warranted inclusion based on subtle interaction signatures.

In \citet{Fuentealba-Fuentes_2025}, we define the close-pair fraction as the number of individual galaxies in close pairs ($N_\mathrm{gm}$) divided by the total number of galaxies. This is also referred to as the galaxy merger fraction \citep[][]{Conselice_2006}, and differs from the more commonly used definition of the merger fraction as the number of pairs ($N_{\mathrm{m}}$) divided by the total number of galaxies. To obtain the galaxy merger fraction for our structural analysis, we use:
\begin{equation}
    F_\mathrm{gm} = \frac{ 2~ N_\mathrm{m} }{N_\mathrm{m} + N_\mathrm{tot}},
\end{equation}
where $N_{m}$ is the number of mergers and $N_{\mathrm{tot}}$ represents the total number of galaxies in the sample. This definition assumes that each interaction involves at least two galaxies and that the companion is not already included in the parent sample. The relation between the two definitions is $N_\mathrm{gm} = 2 N_\mathrm{m}$, for pair-based studies.

\begin{figure*}
    \centering
	\includegraphics[scale=0.28]{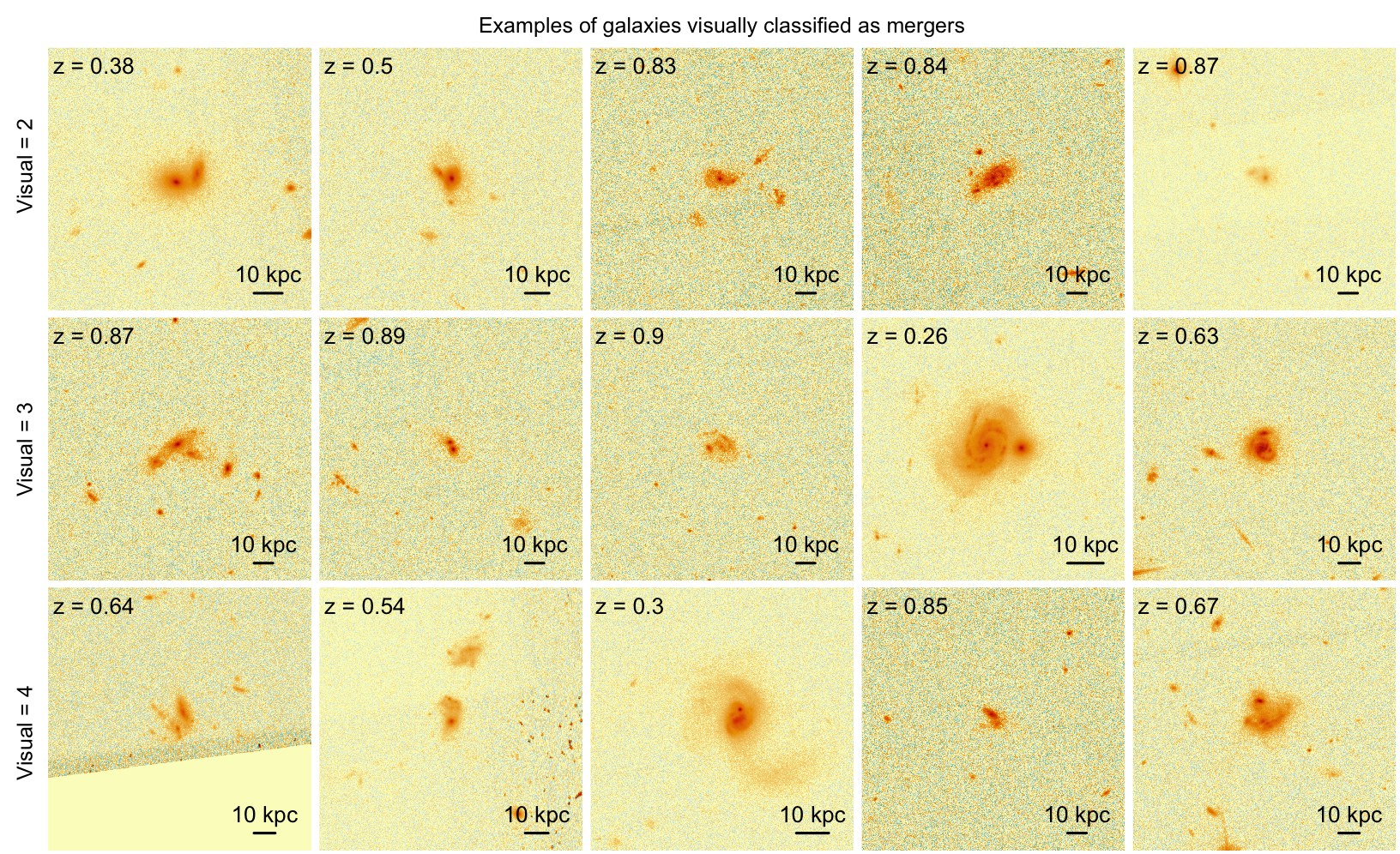}
    \caption{Examples of galaxies visually classified as mergers. Each row represents the consensus of  2, 3, or all 4 classifiers. Galaxies were randomly selected within each category, with their redshifts shown in the corner of each image.}
    \label{fig:visuals}
\end{figure*}

Using this equation, we obtained three different galaxy merger fractions by separating galaxies based on how many of the classifiers identified them as major mergers. The first measurement, $F_\mathrm{vis=4}$, considers only galaxies classified as major mergers by all four classifiers, representing a lower limit for the visual classification. The second estimate, $F_\mathrm{vis\geq2}$, includes galaxies identified as major mergers by two or more classifiers, providing an upper limit for the merger fraction. Finally, the third fraction, $F_\mathrm{vis\geq3}$, counts galaxies classified as major mergers by three or more classifiers (see Figure \ref{fig:visuals} for examples of galaxies classified as mergers by 2, 3, or 4 classifiers).

\begin{figure*}
    \centering
	\includegraphics[scale=0.46]{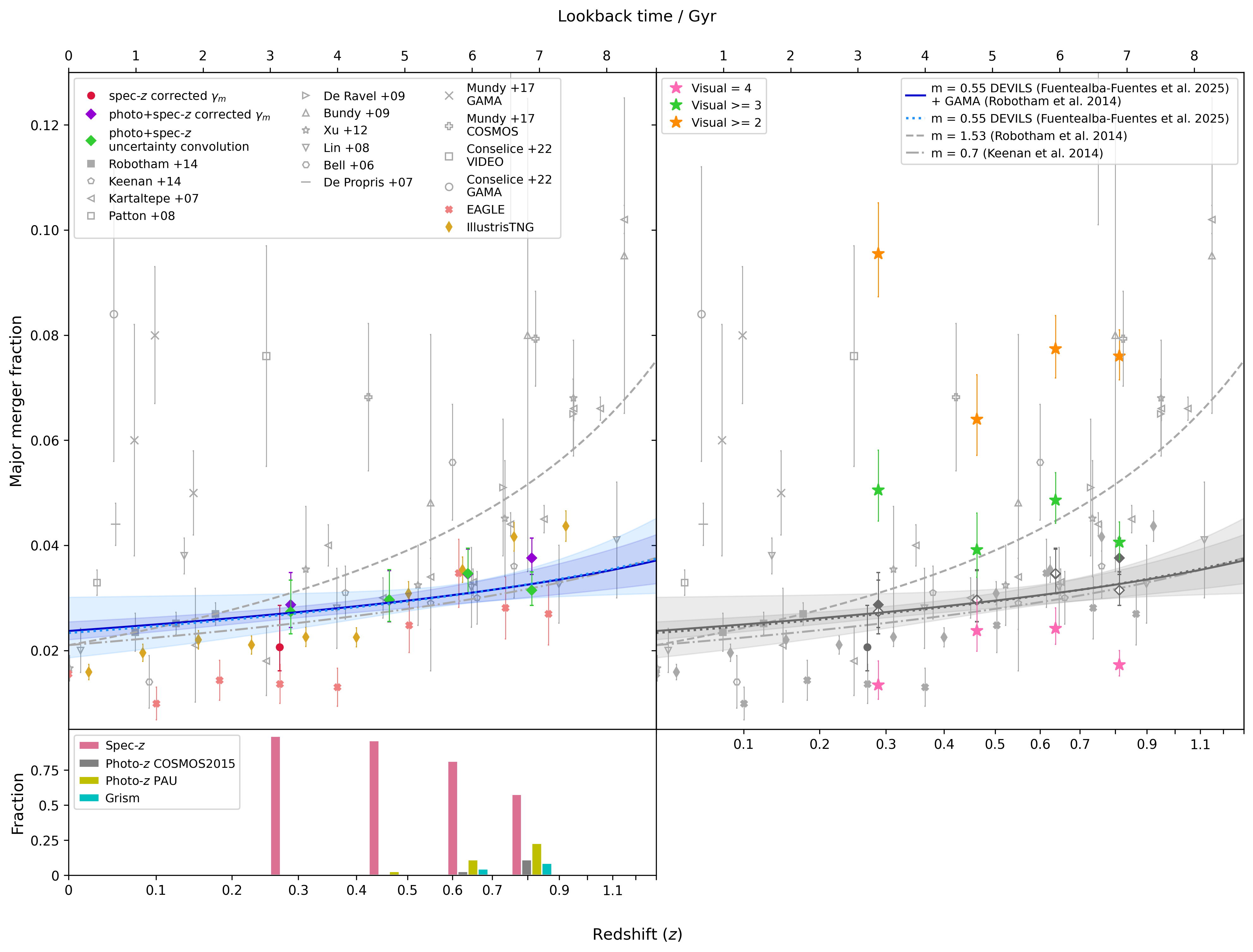}
    \caption{Galaxy merger fractions using the D10 (COSMOS) sample. \textbf{Top-left:} Galaxy merger fractions estimated from close-pair selections, as presented in \citet{Fuentealba-Fuentes_2025} are shown in different colours. Predictions from the Eagle and IllustrisTNG simulations are also shown. Works from the literature based on close-pair samples are shown in grey. \textbf{Top-right:} Galaxy merger fractions estimated from visual classifications of galaxies are presented in different colours. Pink stars indicate galaxies classified as mergers by all four classifiers, green stars those classified by three or more classifiers, and orange stars those classified by two or more classifiers. 
    \textbf{Bottom:} The fraction of spectroscopic, photometric, and grism redshifts in each bin. }
    \label{fig:MergerFractionsVisual}
\end{figure*}

Figure \ref{fig:MergerFractionsVisual} shows the galaxy merger fractions presented in \citet{Fuentealba-Fuentes_2025} as coloured symbols, while the grey symbols represent other close-pair studies from the literature \citep[][]{Bell_2006, De_Propris_2007, Kartaltepe_2007, Patton_2008, Lin_2008, De_Ravel_2009, Bundy_2009, Xu_2012, Robotham_2014, Keenan_2014, Mundy_2017, Conselice_2022} that have been converted into galaxy merger fractions, where required, using $N_\mathrm{gm} = 2 N_\mathrm{m}$.We now also include the estimates from the visual classifications. The uncertainties on these measurements are estimated based on the confidence intervals
on binomial population proportions, as described in \citet{Cameron_2011}. For $F_\mathrm{vis=4}$, shown as pink stars, we obtain lower values than those derived from close-pair selections (dark grey circle and the filled and unfilled dark grey diamonds), and these appear to be consistent with the predictions from the {\sc Eagle} simulations. This better agreement with {\sc EAGLE} is expected, as the visual classification is likely isolating late-stage mergers, while the measurement in the simulation is based on identifying the time at which two galaxies can no longer be separated in 3D space. For $F_\mathrm{vis\geq3}$, the galaxy merger fractions (green stars) are higher and slightly exceed the close-pairs estimates, while $F_\mathrm{vis\geq2}$ yields the highest galaxy merger fractions (orange stars), representing an inclusive upper limit for visually identified mergers. 

Notably, the increasing trend in galaxy merger fractions with redshift, which is seen in the literature as well as in our previous work \citep[e.g.][]{Robotham_2014, Keenan_2014, Fuentealba-Fuentes_2025}, is not recovered in any of the visual merger fractions. Instead, the visually identified fractions exhibit no clear evolution across lookback time, remaining broadly constant with mild fluctuations between bins. This highlights that visually identifying major mergers is highly subjective, as the variation among the four classifiers in this work is substantial, both in the values and in the trends obtained. This will be discussed further in Section \ref{Discussion}.

\begin{figure}
    \centering
	\includegraphics[scale=0.137]{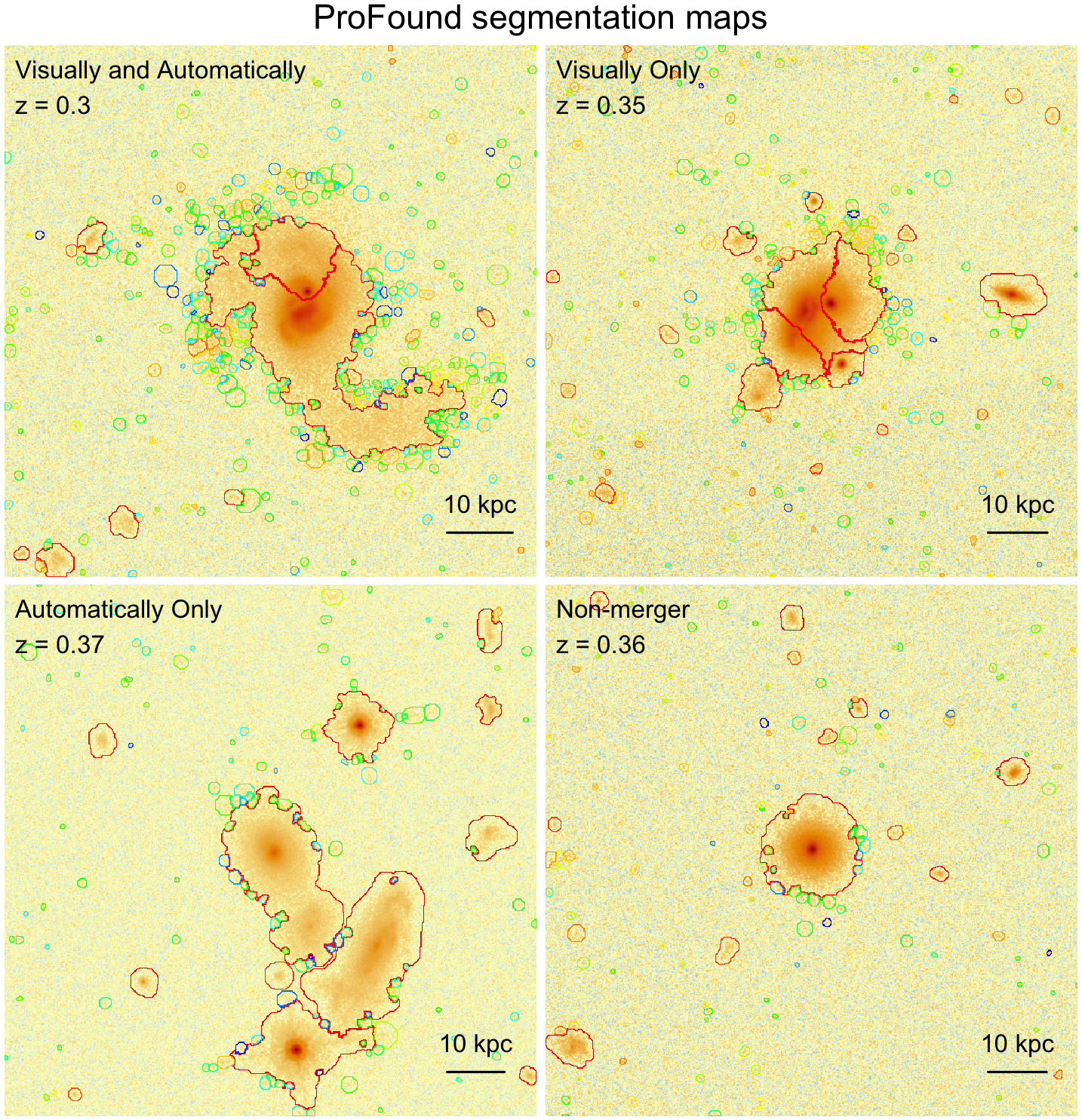}
    \caption{Examples of segmentation maps derived from \texttt{ProFound}. Panels are categorized by merger classification: both visual and automated, visual only, automated only, and non-mergers. Here, "Visual" indicates galaxies identified as mergers by all four classifiers, while no visual confirmation indicates galaxies classified as non-mergers by all.}
    \label{fig:segmap}
\end{figure}
  
\section{Automated classification} \label{Automated}

For an automated classification of mergers, in addition to the HST/ACS cutouts, we require a segmentation map for each galaxy. These maps are essential because they define which pixels belong to each object, allowing us to measure structural parameters consistently and without contamination from neighbouring sources or background noise. We estimate the segmentation maps using the R package \texttt{ProFound} \citep{Robotham_2018}. Unlike traditional circular or elliptical apertures, \texttt{ProFound} is a source-detection tool that creates segments with irregular apertures, better matching the shapes of galaxies commonly seen at higher redshifts, faint isophotal levels, or in crowded fields. It employs segment dilation to grow each segment and estimate pseudo-total fluxes, while watershed deblending prevents overlap, assigning each pixel to a single object. 

While using \texttt{ProFound}, we provide the gain, magnitude zero-point, and pixel scale of the images, as well as the sky value set to 0 since the images are already sky-subtracted. We set redosky = FALSE. A significant challenge in morphological merger studies is ensuring that tidal features and secondary nuclei are not "shredded" into separate segments. To prevent this, we adjusted the \texttt{tolerance} and \texttt{reltol} parameters in \texttt{ProFound}. The \texttt{tolerance} determines how much peak flux an object must have relative to neighbouring objects before being merged; we set it to $5$. The \texttt{reltol} parameter modifies the  \texttt{tolerance} based on the ratio of the segment peak flux to the saddle point flux where it touches a neighbouring segment, raised to the power of \texttt{reltol}. We use \texttt{reltol} = $0.6$, which allows more aggressive merging in the outskirts of galaxies, where the peak flux is typically much larger than the saddle point flux. In the standard DEVILS data processing, these parameters were set to \texttt{tolerance} = $0.8$ and \texttt{reltol} = $0$ (the default value). With our revised configuration, we obtain the segmentation maps used throughout the rest of this work (see Figure \ref{fig:segmap}).

Using these HST/ACS cutouts and \texttt{ProFound} segmentation maps, we estimate five widely used non-parametric statistics to analyse the morphological properties of galaxies: concentration, asymmetry, smoothness, Gini, and $M_{20}$ \citep[][]{Conselice_2003}. We first attempted to calculate all five statistics using \texttt{statmorph}, a Python package that computes non-parametric morphological measurements and performs 2D Sérsic fitting on galaxy images, as described by \citet{Rodriguez-Gomez_2019}. The \texttt{statmorph} package has been implemented in multiple studies to obtain non-parametric morphological diagnostics of galaxies, including the identification of merger signatures \citep[e.g.][]{Guzman-Ortega_2023}, as well as to assess the reliability of these statistics in identifying post-merger systems \citep[][]{Wilkinson_2024}. We analysed the distribution of the galaxy sample across these different parameters, using the visually identified mergers as a reference to evaluate how effectively these parameter spaces separate mergers from non-mergers. Mergers are generally expected to occupy distinct regions of these spaces, typically exhibiting high asymmetry, low concentration, high Gini coefficients, and high $M_{20}$ values, reflecting the strong structural disturbances caused by tidal interactions. For smoothness, the index can be elevated in both merging systems and non-merging star-forming galaxies, as it traces small-scale clumpy structure. Consequently, smoothness is not a reliable merger indicator on its own and is typically used together with asymmetry.

\begin{figure*}
    \centering
	\includegraphics[scale=0.6]{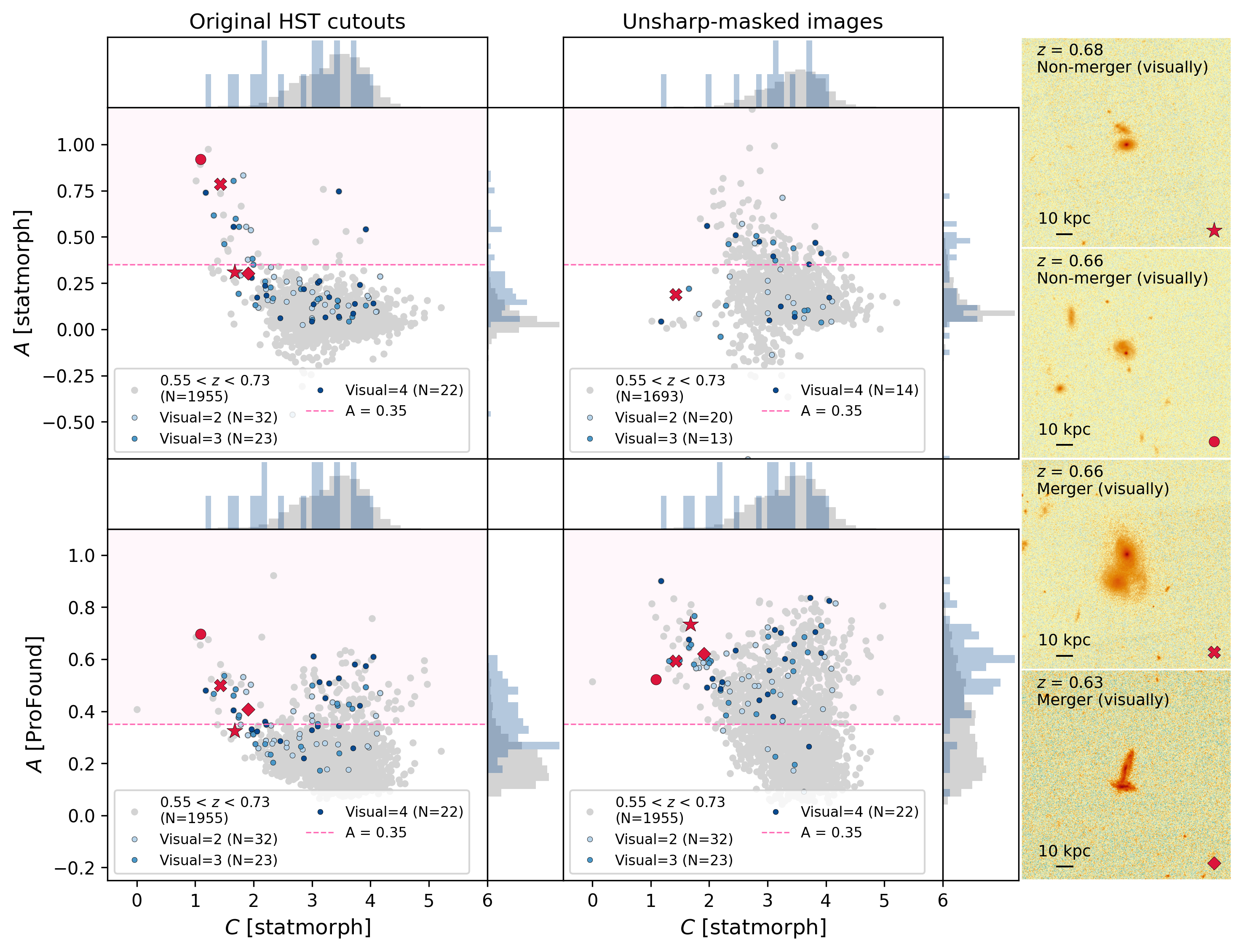}
    \caption{Concentration and asymmetry parameters measured from original and unsharp-masked images for $0.55 < z < 0.73$ galaxies, illustrating the impact of software and images on measured parameters. The concentration index is derived using \texttt{statmorph} in all panels. In the top row, the asymmetry index is calculated using \texttt{statmorph}, while in the bottom row it is measured using \texttt{ProFound}. The left column shows the asymmetry as measured on the original HST/ACS images, and the right columns on the unsharp-masked images. The final column presents examples highlighting contrasting asymmetry values obtained with \texttt{statmorph} and \texttt{ProFound} for visually classified mergers and non-mergers. It is important to note that the asymmetry measurements with \texttt{statmorph} on the unsharp-masked images are done after replacing all negative pixel values of the images with zero, otherwise the sample is significantly reduced, as most galaxies would have flag values $>1$.}

    \label{fig:A_stat_pro}
\end{figure*}

While $C, G$, and $M_{20}$ showed a clearer separation between visually classified mergers and the general galaxy population, the asymmetry values showed a large spread with no clear separation between visually selected mergers and non-mergers, limiting their diagnostic ability. This overlap can be seen in the top-left panel of Figure \ref{fig:A_stat_pro}, which shows galaxies in the third redshift bin ($0.55 < z <0.73$), where asymmetry is measured on the original HST/ACS images, using sources with flag values $\leq 1$, as recommended in the \texttt{statmorph} documentation. In particular, we identified numerous sources with high asymmetry values that lacked visual signatures of interaction, as well as visually confirmed mergers with low asymmetry. 

To address this issue, we adopt the asymmetry measurement derived from \texttt{ProFound} using the \texttt{profoundSegimStats} function. This function provides a summary of various statistics for the individual segments of the image, including the $180$-degree flux asymmetry, ranging from $0$  (perfect symmetry) to $1$ (complete asymmetry), following:
\begin{equation}
    A [\mathrm{ProFound}] = \frac{\sum |I - I_{180}|}{\sum |I + I_{180}|}  ,
\end{equation}
\noindent where $I$ is the original image and $I_{180}$ is the image rotated by 180$^{\circ}$ about the galaxy's center. This definition does not include the subtraction of the background asymmetry ($A_\mathrm{bkg}$) and it normalizes by the sum of the combined flux of the original and rotated images rather than the total flux of the original image alone, which is the standard approach used in common definitions of asymmetry (e.g. Equation~\ref{eq:A_def}) and is the version implemented in \texttt{statmorph}. This \texttt{ProFound} definition is well-suited for our images as they are already sky-subtracted. The \texttt{ProFound} asymmetry more effectively distinguishes visual merging systems from the general non-interacting population (see the bottom-left panel of Figure \ref{fig:A_stat_pro}). The final column of Figure \ref{fig:A_stat_pro} shows examples of contrasting asymmetries between the two asymmetry measures, highlighting the improved separation of the merger population when using the \texttt{ProFound} asymmetry compared to \texttt{statmorph}.

\subsection{Asymmetry and unsharp mask} \label{asymmetry}

While standard asymmetry measurements effectively capture global structural distortions, we found they were less sensitive to the faint, small-scale structures indicative of early-stage interactions or minor disturbances. To enhance these features, we apply an unsharp-masking technique using the \texttt{profoundImDiff} function. We generate a residual image ($I_\mathrm{resid}$) by subtracting a smoothed version of the galaxy cutout ($I_\mathrm{smooth}$) from the original image ($I_\mathrm{orig}$):

\begin{equation}
    I_\mathrm{resid} = I_\mathrm{orig} - I_\mathrm{smooth}. 
    \label{eq:unsharp}
\end{equation}

\begin{figure}
    \centering
    \includegraphics[scale=0.137]{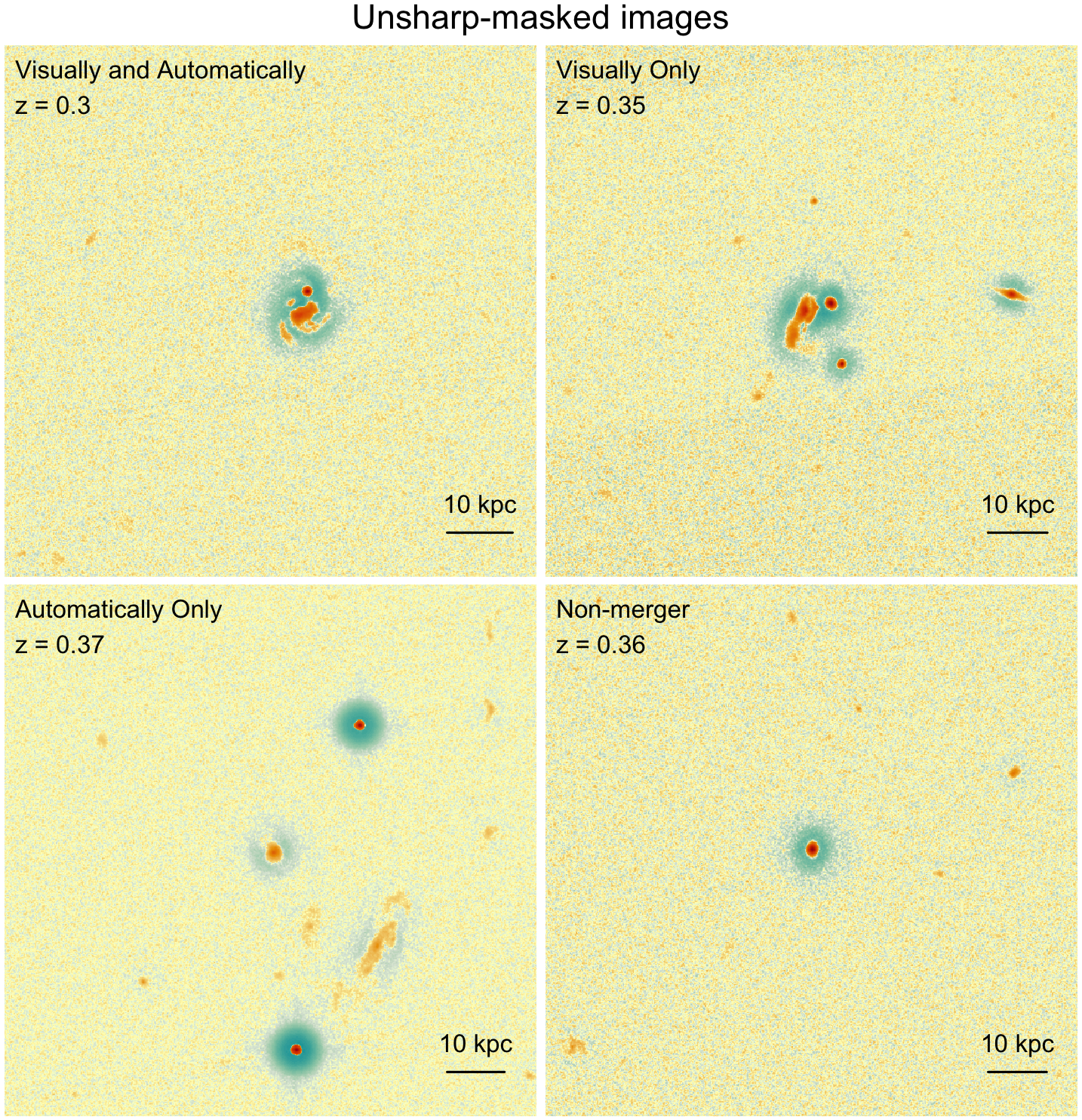}
    \caption{Examples of unsharp-masked images obtained 
    using \texttt{profoundImDiff}. Panels are categorized by merger classification: both visual and automated, visual only, automated only, and non-mergers. Here, "Visual" indicates galaxies identified as mergers by all four classifiers, while no visual confirmation indicates galaxies classified as non-mergers by all.}
    \label{fig:unsharped}
\end{figure}

\noindent The smoothed image is produced by applying a 2D Gaussian blur with a standard deviation of $\sigma = 10$ pixels. We chose this value after testing seven different scales ($\sigma = 1, 3, 5, 10, 30, 50$, and $100$), with $\sigma = 10$, providing the best balance between the improvement in signal-to-noise ratio and the preservation of the original morphology. By subtracting this smoothed version from the original image, we produce a residual that isolates high-frequency, non-smooth structures such as spiral arms, clumps, or tidal features (see Figure \ref{fig:unsharped}) from the smooth underlying light distribution of the galaxy. We then re-measured the asymmetry using these unsharp-masked images and the same segmentation maps as before. We found that this approach significantly improved the ability of the asymmetry statistic to separate merging systems from non-interacting galaxies, specifically as defined by our visual classification. This can be seen in Figure \ref{fig:A_stat_pro} bottom-middle panel, where the visual merger sample is more clearly isolated from the parent sample. In contrast, the top-middle panel shows the asymmetry measured with \texttt{statmorph} on the same unsharp-masked images, considering only galaxies with flag values $\leq 1$. Applying this quality cut reduces the sample considerably more than when using \texttt{ProFound}. These measurements are performed after replacing all negative pixel values in the images with zero; otherwise, the sample would be further reduced, as most galaxies would have flag values $> 1$.

\subsection{Merger selection criteria}

A common automated selection criteria used in the literature is the one set by \citet{Conselice_2003} and \citet{Lotz_2008b} that includes the asymmetry, smoothness, Gini, and $M_{20}$ coefficient in two equations:

\begin{align}
   f(G,M_\mathrm{20}) &= G + 0.14\, M_\mathrm{20} > 0.33,
   \label{eq:G-M20_lit} \\[6pt] 
   A &\geq 0.35 \, \textrm{and } A > S,
   \label{eq:A_lit}
\end{align}

\noindent These equations were derived from HST observations and have been validated up to $z \sim 1.2$. 

\begin{figure}
    \centering
	\includegraphics[scale=0.65]{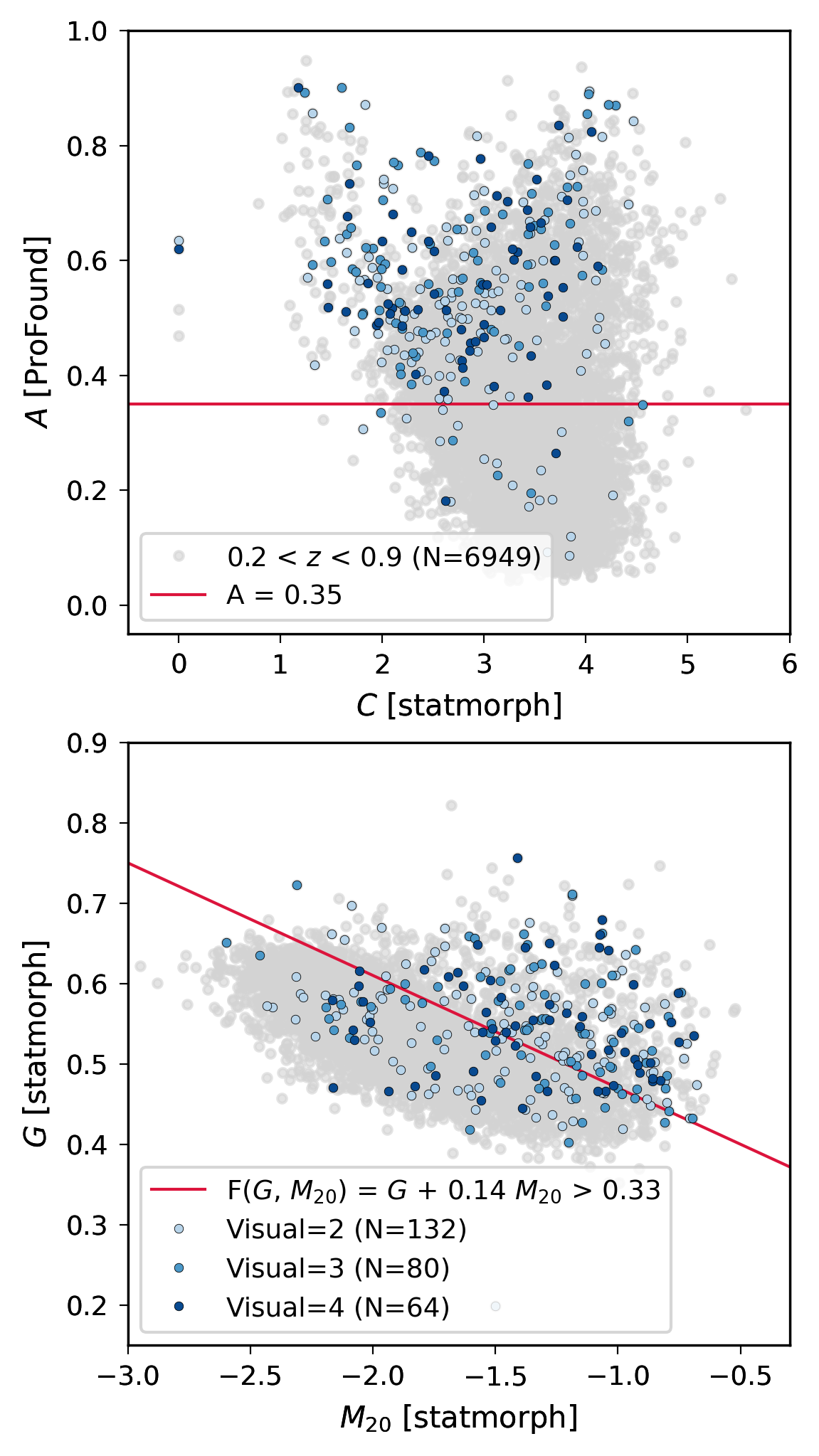} 
    \caption{Non-parametric merger selection. The top row shows the asymmetry using the unsharp-masked images produced using \texttt{profoundImDiff} vs. the concentration from \texttt{statmorph}. The bottom row shows the Gini vs. $M_{20}$ space, both from \texttt{statmorph}. Grey points represent the full galaxy sample within the full redshift range. Galaxies visually classified as mergers are shown as blue points, with light to dark shades indicating agreement by two, three, and all four classifiers, respectively. The selection thresholds commonly adopted in the literature, defined by Equations~\ref{eq:G-M20_lit} and~\ref{eq:A_lit}, are shown as red lines.}
    \label{fig:Threshold_original}
\end{figure}

Although the value of $C$ is not always included as a determinant for the merger selection in standard equations, asymmetry is typically plotted against the concentration index. In this $A$--$C$ parameter space, including the $A > S$ condition, it is expected that mergers can be identified by quantifying how disturbed the light distribution is relative to the galaxy's concentration. Usually, these thresholds (regardless of whether $C$ is explicitly included) identify much later stages of a merger, as galaxies must be highly disturbed to be recognized as merging systems.
In contrast, the $G$--$M_{20}$ parameter space can identify wider stages of the merger, as it is particularly sensitive to the presence of multiple bright nuclei and the spatial distribution of the brightest pixels. Even if the overall light distribution of the galaxy remains relatively symmetric, $G$--$M_{20}$ can still detect early signs of the merger.

\begin{figure*}
    \centering
	\includegraphics[scale=0.32]{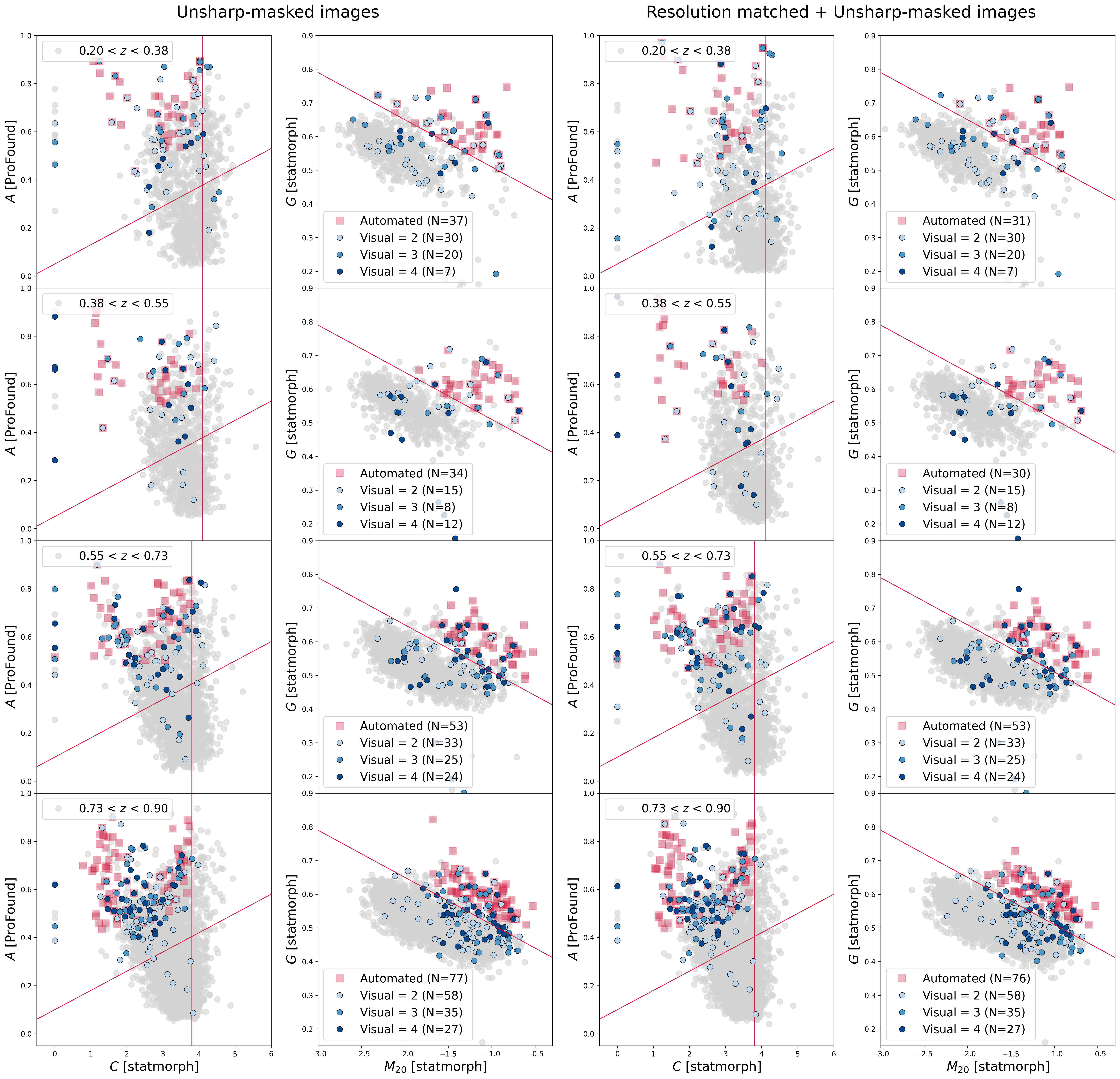}
    \caption{Non-parametric merger selection. The first two columns on the left show the asymmetry vs. concentration and Gini vs. $M_{20}$ spaces, respectively, using the unsharp-masked images produced using \texttt{profoundImDiff}. The two rightmost columns show the same parameter spaces measured on images that were first downgraded in physical resolution and then unsharp-masked using \texttt{profoundImBlur} and \texttt{profoundImDiff}, respectively. In all panels, grey points represent the full galaxy sample within the redshift range of each row. Galaxies visually classified as mergers are shown as blue points, with light to dark shades indicating agreement by two, three, and all four classifiers, respectively. Red squares correspond to galaxies that were classified as mergers by the automated non-parametric selection based on $A, C,$ $S$, $G$, and $M_{20}$. The corresponding selection thresholds are defined by Equations~\ref{eq:mergerG-M20}, \ref{eq:mergerC}, and \ref{eq:mergerA}, and are shown as red lines.}
    \label{fig:Autoclass}
\end{figure*}

However, we found that the distribution of our sources in these parameter spaces was not efficient at separating merging galaxies, as these thresholds did not identify a number of galaxies that were visually classified as mergers (see Figure \ref{fig:Threshold_original}). Therefore, we tested different thresholds for the spaces $A$--$C$ and $G$--$M_{20}$ and defined the following selection criteria for automated classification:

\begin{align}
f(G, M_{20}) &= 
\begin{cases}
G + 0.14\,M_{20} > 0.37,  & \text{if } 0.2 < z < 0.55, \\
G + 0.14\,M_{20} > 0.4,  & \text{if } 0.55 < z < 0.9,
\end{cases}
\label{eq:mergerG-M20} \\[6pt]
C &\geq 
\begin{cases}
4.1, & \text{if } 0.2 < z < 0.55, \\
3.8, & \text{if } 0.55 < z < 0.9,
\end{cases}
\label{eq:mergerC} \\[6pt]
A &\geq 0.08\,C + 0.25 \, \textrm{and } A > S,
\label{eq:mergerA}
\end{align}

The key differences between the two sets of equations are that the asymmetry threshold now depends on the galaxy’s concentration, and we also define a concentration limit with two possible values depending on the redshift of the sources: one for lower-$z$ galaxies ($0.2 < z < 0.55$) and another for higher-$z$ sources ($0.55 < z < 0.9$). Similarly, for the $G$--$M_{20}$ relation, we have modified the intercept coefficient, again providing two options based on the same lower-$z$ and higher-$z$ bins. These selections were defined by eye using the visual samples, as they were found to provide the most effective separation of galaxies visually classified as mergers. This can be seen in Figure \ref{fig:Autoclass}, where the full sample is divided into the same four redshift bins used by \citet{Fuentealba-Fuentes_2025}, with each row corresponding to one bin. The two left columns show how the automated classifier separates mergers from non-mergers. Red squares indicate galaxies classified as mergers by the automated method, while blue points show visually classified mergers. The shades of blue reflect the level of consensus among the classifiers, with darker shades corresponding to higher agreement (all four have identified the galaxies as mergers).

\begin{figure}
    \centering
	\includegraphics[scale=0.54]{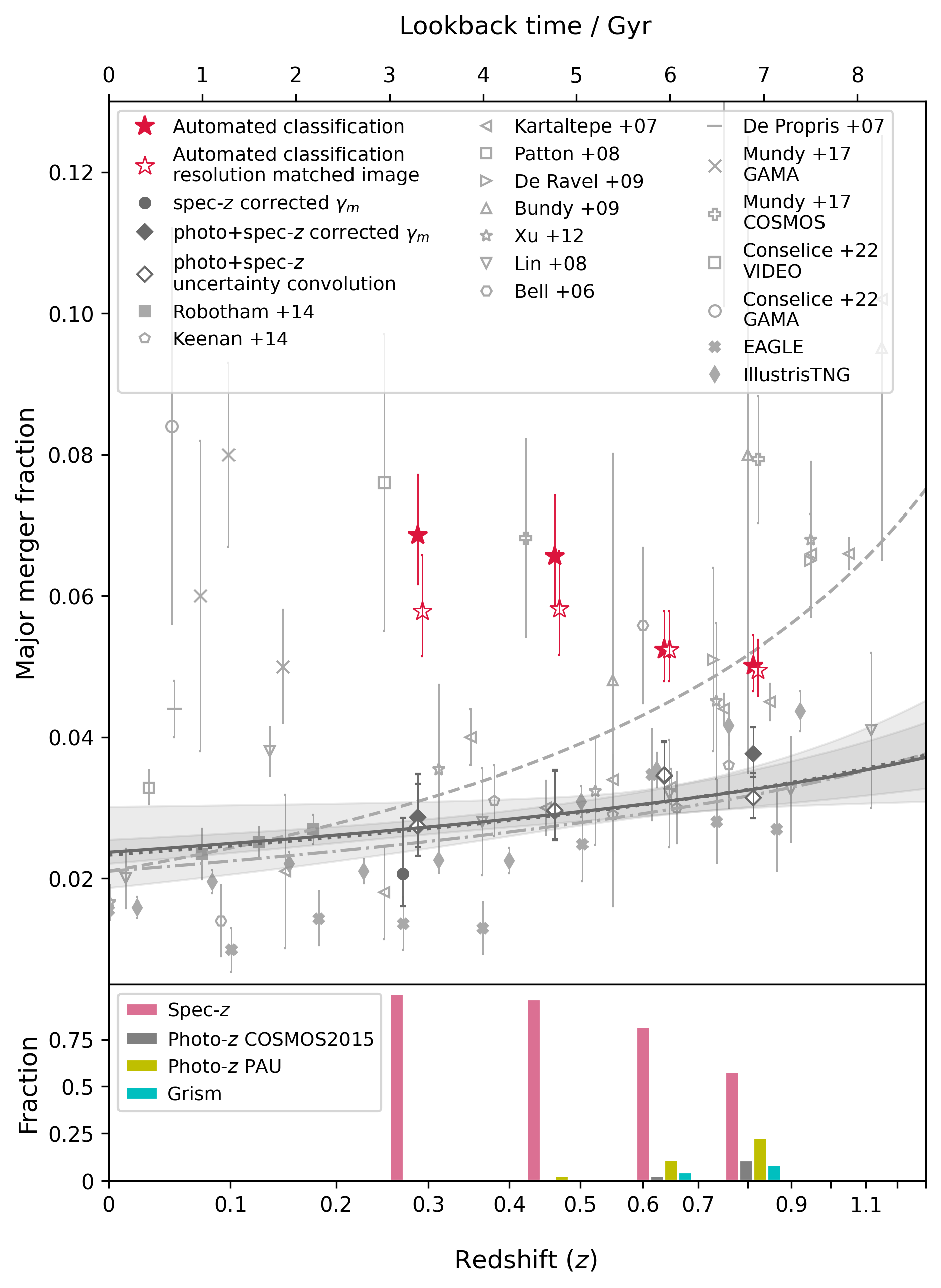}
    \caption{Galaxy merger fractions using the D10 (COSMOS) sample. galaxy merger fractions estimated from close-pair selections, as presented in \citet{Fuentealba-Fuentes_2025}, are shown as grey circles and filled and unfilled grey diamonds. \textbf{Top:} Galaxy merger fractions estimated from automated classifications using asymmetry from \texttt{ProFound}, and concentration, Gini, and $M_{20}$ from \texttt{statmorph}. Filled red stars correspond to classifications performed on the unsharp-masked images, while unfilled red stars show results obtained from the physical resolution matched $+$ unsharp-masked images.
    \textbf{Bottom:} The fraction of spectroscopic, photometric, and grism redshifts in each bin. }
    \label{fig:MergerFractionsAuto}
\end{figure}

The galaxy merger fractions obtained using this selection criteria and measured on the unsharp-masked images (see Section \ref{asymmetry}) are presented as filled red stars in Figure \ref{fig:MergerFractionsAuto}. The uncertainties on these estimates are calculated in the same way as those for the visual fractions, based on the confidence intervals of binomial population proportions, as described in \citet{Cameron_2011}. We find that these measurements are significantly higher than the close-pair fractions reported in \citet{Fuentealba-Fuentes_2025} and fall within the range of values obtained from our visual classifications. The galaxy merger fractions decrease with increasing redshift, likely reflecting both the high spatial resolution of our HST/ACS mosaic and cosmological effects. As redshift increases, the fixed pixel scale samples progressively larger physical areas, effectively lowering the physical resolution. Combined with surface brightness dimming, this likely causes the automated diagnostics to miss faint tidal features or subtle asymmetries that are more easily detected at lower redshifts. To quantify the extent of this resolution bias, we systematically degraded the resolution of the HST/ACS images, as described in the following section.

\subsection{Physical resolution matched images}

To determine if the performance of our automated classification is biased by the varying physical resolution across the redshift range, we performed a degradation test on our cutouts. Using \texttt{profoundImBlur}, we produced a smoothed version of each image by applying a Gaussian kernel with a standard deviation ($\sigma$). This $\sigma$ was specifically calculated to match the physical resolution of a galaxy at our highest redshift limit, $z = 0.9$, effectively simulating how low-redshift systems would appear if observed at the edge of our volume.

\begin{figure}
   \centering
	\includegraphics[scale=0.137]{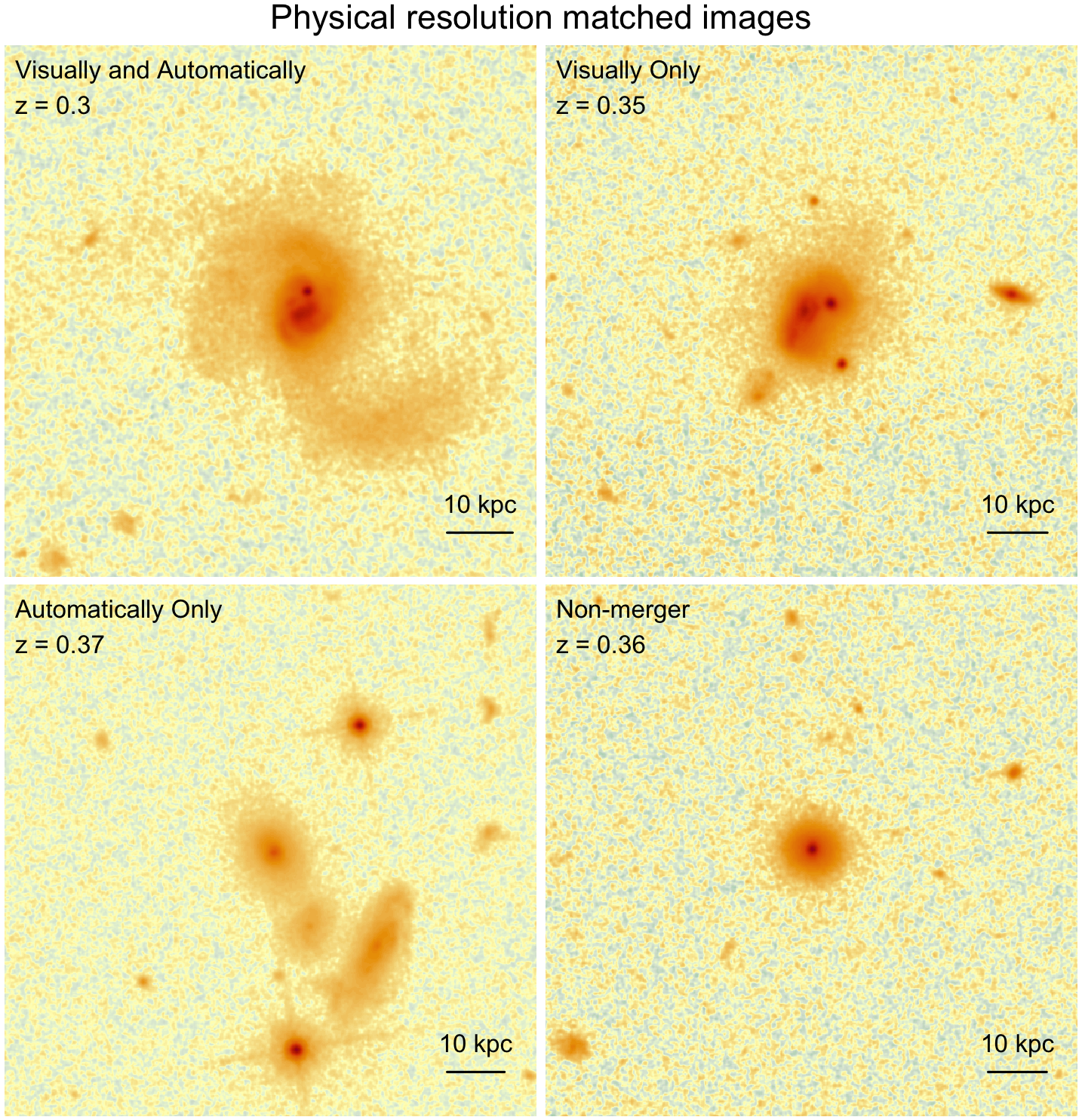}
    \caption{Examples of physical resolution matched images derived from \texttt{profoundImblur}. Panels are categorized by merger classification: both visual and automated, visual only, automated only, and non-mergers. Here, "Visual" indicates galaxies identified as mergers by all four classifiers, while no visual confirmation indicates galaxies classified as non-mergers by all.}
    \label{fig:resolution}
\end{figure}

The resulting lower-resolution images (see Figure \ref{fig:resolution}) are then processed as before to create unsharp masks using \texttt{profoundImDiff}, allowing us to estimate the non-parametric coefficients ($C$, $A$, $S$, $G$, and $M_{20}$) and identify merging galaxies using the same selection thresholds defined previously (see the two right columns of Figure \ref{fig:Autoclass}). 

This comparison allows us to quantify how the higher spatial resolution available at lower redshifts may influence the observed galaxy merger fractions, ensuring that any measured evolution (or lack thereof) is not merely a consequence of changing physical resolution with lookback time. The galaxy merger fractions obtained from these physical resolution matched $+$ unsharp-masked images are shown as unfilled red stars in Figure \ref{fig:MergerFractionsAuto}. We find that in the two lowest redshift bins, the galaxy merger fractions derived from physical resolution matched $+$ unsharp-masked images are slightly lower than their original measurements (see filled vs empty stars in Figure \ref{fig:MergerFractionsAuto}), suggesting that higher resolution allows the detection of more subtle merger signatures. Conversely, the results in the two highest redshift bins remain mostly consistent across both approaches, which is expected as the original images at $z \sim 0.9$ already possess a physical resolution similar to our blurring kernel. 

\section{Discussion} \label{Discussion}

\subsection{Galaxy merger fractions from visual and automated morphological classifications}

The recovery of visually identified galaxies by automated classification remains largely stable across the different visual classification categories (unanimous consensus, three or more, and two or more classifiers), with a slight improvement observed as classifier consensus increases. On the unsharp-masked images, we find that between $\sim 15 - 30\%$ of galaxies visually classified as mergers are recovered by both the $G$--$M_{20}$ and $A$--$C$ criteria, with the exception of the first redshift bin ($0.2 < z < 0.38$), where none of the galaxies with unanimous consensus are recovered by the automated pipeline. 

When applying the same criteria to the physical resolution matched $+$ unsharp-masked images, this recovery rate is generally maintained. However, we observe a decrease in these rates for the second redshift bin ($0.38 < z < 0.55$), with a slight increase for the first bin, reaching the $\sim 30\%$ upper limit for the sample with unanimous visual agreement. Overall, the overlap between visually and automatically classified samples is relatively low.

The discrepancy between visual and automated classifications likely arises from the inherent limitations of each method in distinguishing true dynamical interactions from "contaminant" features. Both approaches are affected by decreasing spatial resolution and cosmological surface brightness dimming ($(1+z)^4$). Automated non-parametric statistics can be affected by contamination from non-tidal features, particularly in clumpy, star-forming galaxies and dusty systems, potentially resulting in a higher false-positive rate. Conversely, visual classification is limited by the eye's inability to detect low-surface-brightness structures at high redshift and remains inherently subjective, as it depends on the individual classifier's judgment.

Furthermore, the rest-frame wavelength of HST imaging shifts to bluer light at higher-$z$, tracing the rest-frame ultraviolet (UV), which can complicate merger identification because it highlights star-forming regions that may be mistaken for interaction-induced disturbances. Observations with the James Webb Space Telescope (JWST) could mitigate these effects by probing the rest-frame near-infrared (NIR) at high redshift. Ideally, combining HST and JWST data to produce stellar-mass-weighted images would provide a more consistent view of the galaxy structure.

\subsubsection{Comparison of the $G$--$M_{20}$ and $A$--$C$ spaces as automated diagnostics}

We evaluate the performance of our automated diagnostics by comparing them against the visual merger classifications. In Figure \ref{fig:Autoclass}, grey points represent the full sample, while blue points indicate visual mergers (with darker shades denoting higher classifier consensus). Red squares represent galaxies classified as mergers by both $G$--$M_{20}$ and $A$--$C$ selection criteria simultaneously.

The $A$--$C$ space defined by Equations \ref{eq:mergerC} and \ref{eq:mergerA} shows the closest agreement with the visual merger sample (blue points). When measured on unsharp-masked images, this selection remains robust regardless of whether the images are in their original state or have been downgraded in physical resolution to simulate higher redshifts.

Conversely, the $G$--$M_{20}$ space defined by Equation \ref{eq:mergerG-M20} struggles to isolate visually identified mergers without including a significant fraction of the general galaxy population. To avoid high levels of contamination, our thresholds were defined strictly; while this ensures that the automated detections are high-confidence candidates, it results in many visually identified mergers being excluded from the selection.

It is important to note that $CAS$ diagnostics and $G$--$M_{20}$ statistics are sensitive to different stellar mass ratios of merging systems. While $CAS$ parameters are most effective at detecting major mergers with mass ratios $\geq 1:4$, the $G$--$M_{20}$ space can identify a broader range of mergers, extending down to mass ratios of $\sim 1:10$ \citep[][]{Conselice_2006, Lotz_2010}. This wider sensitivity may help explain the larger number of potential mergers identified by the $G$--$M_{20}$ selection, albeit at the cost of higher contamination from minor mergers.

\begin{figure}
    \centering
	\includegraphics[scale=0.345]{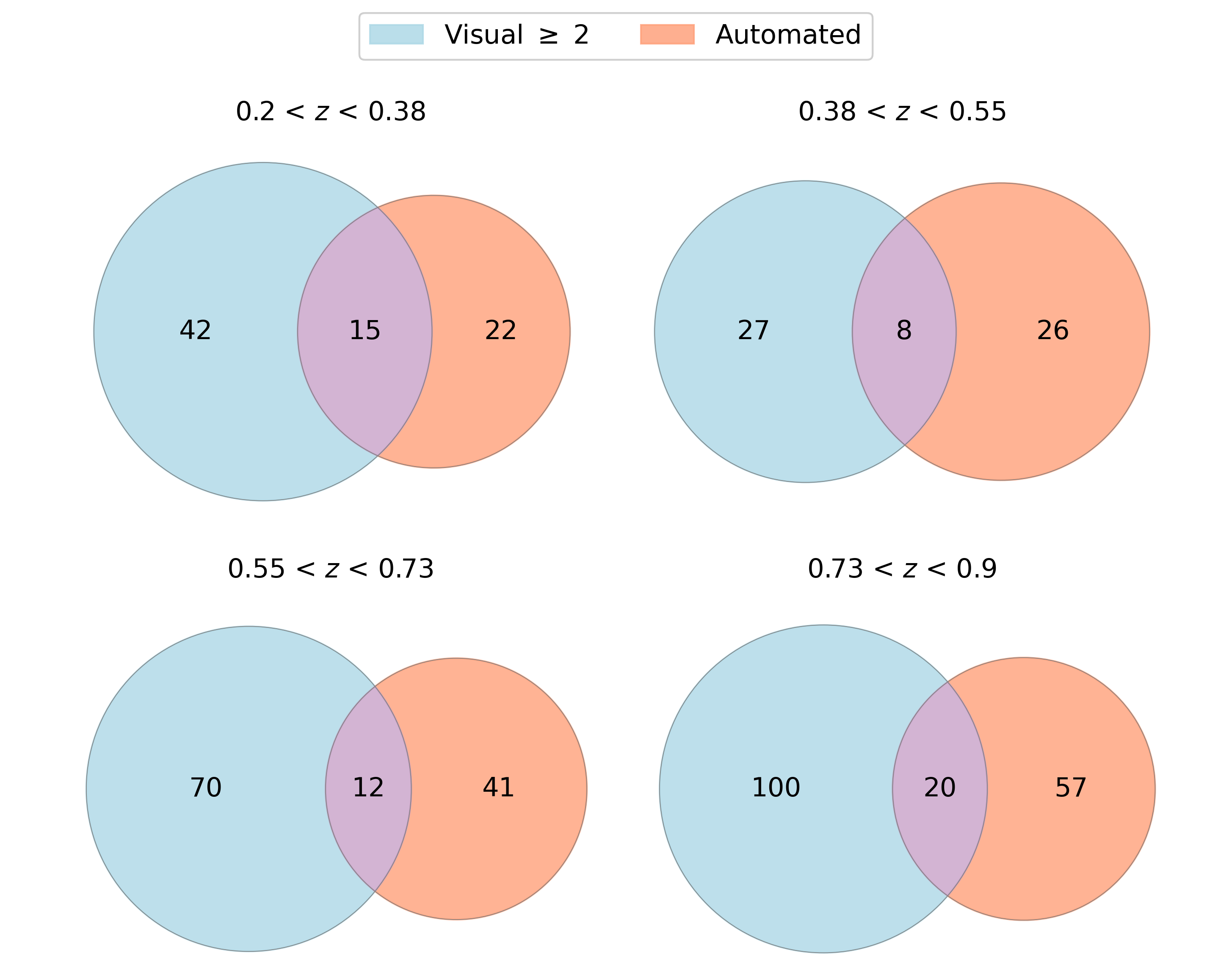}
    \caption{Venn diagrams showing the overlap between the automated sample and our broader visual sample (defined as galaxies with two or more visual merger flags) for each redshift bin.}
    \label{fig:venn}
\end{figure}

\begin{figure*}
    \centering
	\includegraphics[scale=0.5]{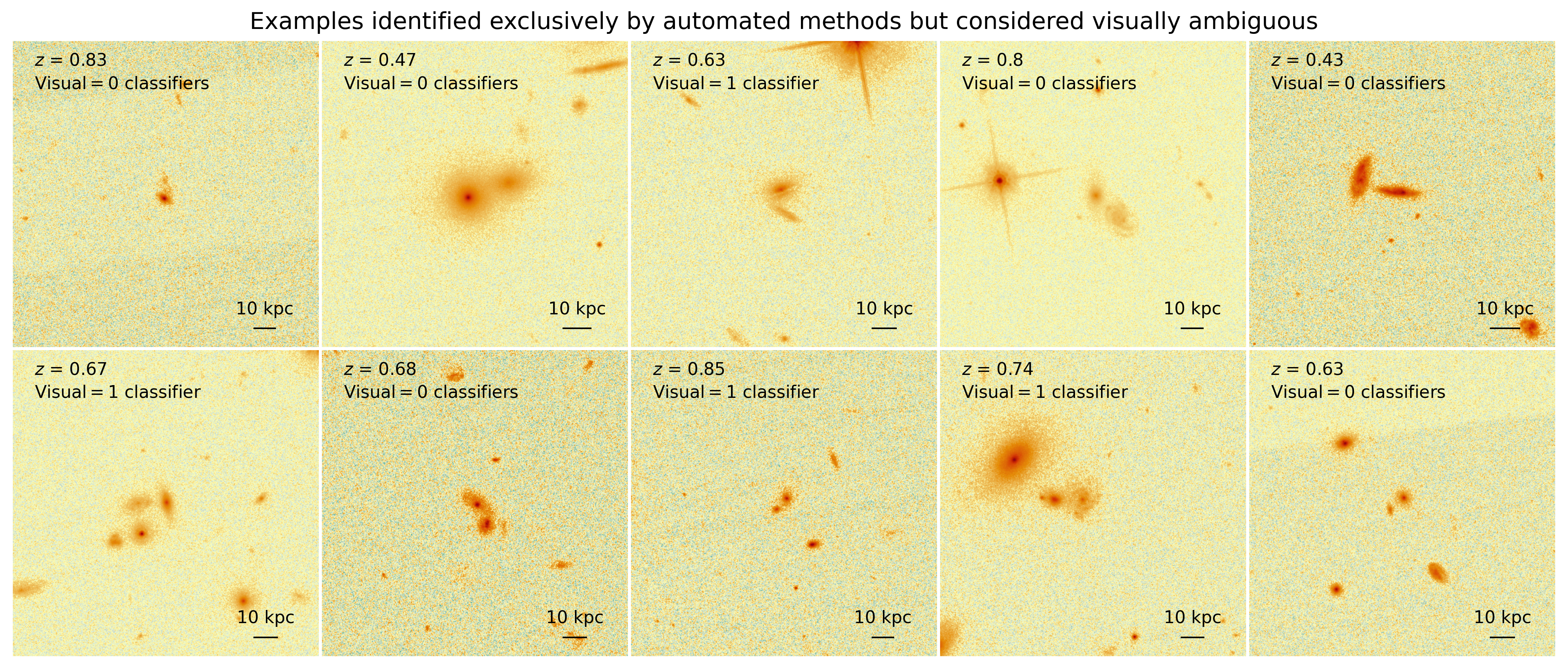}
    \caption{Examples of galaxies classified as mergers by automated methods only, but considered morphologically ambiguous by the classifiers after visual re-inspection. The original number of classifiers identifying each system as a merger is shown in each panel, along with the source redshift.}
    \label{fig:ambiguous}
\end{figure*}

The intersection of both methods (red squares) provides the most conservative census, ensuring that any identified evolution is driven by high-confidence morphological disturbances rather than noise or contamination. However, the overlap between this automated sample and our broader visual sample, which includes galaxies with two or more visual merger flags (all blue points), is relatively low (see Figure \ref{fig:venn}). To understand this discrepancy, galaxies in the automated sample but absent from the visual merger list were re-inspected. In most cases, the classifiers agreed that these systems were morphologically ambiguous (see Figure \ref{fig:ambiguous}), making their merger status debatable even upon close inspection. In a minority of instances, the automated classification likely failed due to segmentation maps that incorrectly blended neighboring sources or mischaracterized the boundaries of complex systems.

\subsection{Close-pair and morphological merger identification}

We find that the galaxy merger fractions derived from most of our morphological techniques are systematically higher than the close-pair fractions reported in \citet{Fuentealba-Fuentes_2025}. Similar results were found in \citet{Conselice_2009} for the COSMOS field where galaxy merger fractions were estimated using the $CAS$ diagnostics and compared to pair-studies from \citet{Kartaltepe_2007}. Their analysis demonstrated that while the galaxy merger fraction remains relatively low at $z < 1$ for massive galaxies, morphological diagnostics consistently recover higher fractions than pair-based selections. The ratio between these two measurements, defined as $\kappa' = f_\mathrm{CAS}/f_\mathrm{pair}$, yielded values from $1.5 - 3.5$, with the caveat that the two studies used different stellar mass cuts. In our analysis, we obtain $\kappa' \sim 1.8 - 2$ when comparing the close-pair measurements from \citet{Fuentealba-Fuentes_2025} with the automated morphological fractions presented here. This is consistent with the range reported by \citet{Conselice_2009}, despite the different merger selection criteria used in this work and the fact that we adopt comparable mass selections for both methods.

This discrepancy is expected because the two methods probe different merger stages with distinct observability timescales. By construction, close-pair selection identifies systems prior to coalescence, whereas morphological diagnostics are sensitive to disturbed structures both before and after coalescence. The two methods therefore trace different phases of the same physical process. Our results suggest that the effective observability timescales associated with morphological disturbances are longer than those of close-pair selections in our sample, given that morphological merger fractions are systematically higher than the corresponding pair fractions. This interpretation is consistent with the results of \citet{Conselice_2009}. We emphasize that this comparison focuses on galaxy merger fractions rather than merger rates, which are a more physically descriptive quantity because they incorporate the merger observability timescale and account for the differing identification windows of each method.

We note that the observability timescale for close-pairs ($\tau_{\mathrm{pair}}$) depends strongly on the adopted projected separation limits ($r_{\mathrm{sep}}$) and the associated dynamical friction timescales. In our comparison with \citet{Fuentealba-Fuentes_2025}, we adopted a tight constrain of $r_{\mathrm{sep}} < 20$ $h^{-1}$ kpc, corresponding to an observability timescale of $\tau_{\mathrm{pair}} \sim 0.6$--$1$ Gyr based on the merger timescales derived by \citet{KandW_2008}, \citet{Snyder_2017}, and \citet{Conselice_2022}. In contrast, morphological disturbances, particularly those identified via $G$–$M_{20}$, probe a shorter observability window \citep[$\tau_{\mathrm{obs}} \sim 0.2$–$0.4$ Gyr,][]{Lotz_2010}. Furthermore, for the asymmetry index, it is known that at $z < 1.5$, the observability timescale varies significantly with the gas fraction of the interaction \citep[][]{Lotz_2010, Lotz_2011}.

It is also worth noting that, within the redshift range of our study ($0.2 < z < 0.9$), galaxy merger fractions are expected to remain relatively low for the high-mass regime analysed here ($10^{10.57}\,M_\odot < M_\star < 10^{11.47}\,M_\odot$), especially when compared to low-mass populations which typically exhibit higher fractions \citep[e.g.][]{Conselice_2006}. 

Although both visual and automated morphological fractions exhibit a decreasing trend with redshift, likely due to the resolution and dimming effects, the absolute values remain higher than those of close-pairs across the sampled redshift range. This offset suggests a physical progression: galaxies identified as morphologically disturbed at a given redshift $z$ would probably have been observed as close-pair systems at an earlier epoch ($z$ + $\Delta{z}$). In this framework, the morphological merger fraction may trace the close-pair population from a higher redshift, when the galaxy merger fraction was higher, shifted forward in time by the duration of the transition from the close-pair phase to the later, morphologically disturbed stages. This effect could help explain why morphological diagnostics yield higher galaxy merger fractions than pair counts within the same redshift bin.

A key challenge in directly comparing the two merger results is that close-pairs are selected using stellar mass ratios $\geq 1:3$ (some studies use ratios $\geq 1:4$), whereas morphological classifications do not incorporate stellar mass information. In our analysis, major mergers are identified purely based on the strength of post merger asymmetries. This introduces a disparity in selection depth. While $CAS$ parameters are typically sensitive to major mergers with ratios $\geq 1:4$, $G$--$M_{20}$ diagnostics can also detect weaker interactions, down to mass ratios of $\sim 1:10$. Consequently, the automated morphological sample may capture a broader range of disturbances than the mass-constrained close-pair selection. In addition, a galaxy’s appearance during a merger is highly sensitive to various factors, such as orbital configuration (e.g. the collision angle), encounter velocity, and gas fraction \citep[][]{Lotz_2011}. As a result, some minor mergers may exhibit strong morphological disturbances, while certain major mergers may show comparatively weak signatures. This sensitivity implies that morphological selections may be biased toward highly disturbed systems, potentially including minor mergers, thereby probing a fundamentally different population from mass-ratio-selected close pairs. A major advantage of morphological techniques is that they can be applied to large imaging surveys without spectroscopic follow-up, whereas close-pair identification requires high spectroscopic completeness, which is observationally expensive. Nevertheless, morphological diagnostics remain strongly dependent on image depth and resolution.

Although both techniques are still potentially contaminated by non-mergers, such as chance projections for pairs or clumpy star-formation for morphology, the mass-ratio selection in our previous work provides a more physically motivated definition of what constitutes a major interaction. This robustness is further enhanced by the use of a narrow velocity window ($\Delta v$) for the spectroscopic sample in \citet{Fuentealba-Fuentes_2025}, which significantly minimizes the impact of chance projections. For the broader sample, which includes high-quality spectroscopic, photometric, and grism redshifts within the DEVILS survey, the methodology analyzes the full probability distribution of redshifts ($P(z)$). This approach effectively accounts for inherent uncertainties in photo-$z$ data, further mitigating the influence of chance projections that typically bias pair fractions at high redshift.

Finally, we draw attention to the fact that our visual and automated galaxy merger fractions are broadly consistent with results in the literature based on close-pair fractions, which are primarily based on photometric sources \citep[e.g.][]{Bell_2006, Kartaltepe_2007, Bundy_2009, Mundy_2017, Conselice_2022}. This highlights the challenge of measuring galaxy merger fractions: differences do not only arise from the method used to identify interactions but are also strongly influenced by sample selection, completeness, and other observational biases. These factors help explain the wide range of values reported in the literature, as well as the differing trends in galaxy merger fractions across lookback time \citep[e.g.][]{Robotham_2014, Keenan_2014, Fuentealba-Fuentes_2025}.

\section{Conclusion} \label{Conclusion}

In this work, we use DEVILS data from the D10 (COSMOS) field to estimate galaxy merger fractions over the redshift range $0.2 < z < 0.9$. We use a morphological approach that includes visual classifications and automated non-parametric diagnostics ($C, A,$ $S,$ $G$, and $M_{20}$). To enhance the detection of faint, low-surface-brightness features, we apply unsharp-masking to high-resolution HST imaging, significantly improving the sensitivity of our asymmetry measurements. We compare these results with our previous estimations presented in \citet{Fuentealba-Fuentes_2025} based on close-pair selections. We summarise our main findings as follows:

\begin{enumerate}[label=\Roman*., wide=0pt, leftmargin=0pt, itemindent=2em]

    \item We have defined new thresholds for the automated classification of galaxy mergers using non-parametric diagnostics ($C, G,$ $S,$ and $ M_{20}$ from \texttt{statmorph}, and $A$ from \texttt{ProFound}). These limits were optimized to recover the visually classified merger sample. These automated selection criteria are presented in Equations \ref{eq:mergerG-M20}, \ref{eq:mergerC}, and \ref{eq:mergerA}.

    \item We find that automated diagnostics can recover up to $\sim 30\%$ of visual mergers, representing a low overlap between both samples. This highlights a fundamental trade-off: automated methods offer higher completeness but are more susceptible to contamination from clumpy, star-forming regions, whereas visual classification tends to maintain higher purity but is limited by human bias and sensitivity, especially at higher redshifts or for low-surface-brightness features.

    \item Our morphological galaxy merger fractions are systematically higher than the close-pair fractions reported in \citet{Fuentealba-Fuentes_2025}. While our results show a decreasing trend with redshift, likely due to surface-brightness dimming and resolution effects, they remain consistently above the close-pair fractions. This offset could be partially explained by the fact that morphologically disturbed galaxies at a certain redshift are the later-stage descendants of close-pairs from an earlier epoch. Consequently, the morphological merger fraction reflects the pair-count population from a higher redshift where the intrinsic galaxy merger fraction was higher.

    \item Our results demonstrate that combining multiple identification techniques is essential for a complete merger census. However, relying on visual or automated morphological identification in a single band alone is likely insufficient, as a single wavelength may not capture the full diversity of galactic structures. Similarly, highly complete spectroscopic samples play an important role in close-pair studies by minimizing chance projection effects and enabling more robust merger identification. Even when accounting for these effects, a direct comparison between morphological and pair-based techniques remains challenging because they are sensitive to distinct stages of the merging process with different observability timescales. Assuming comparable intrinsic merger rates, the systematically higher morphological merger fractions measured in our sample may reflect differences in the effective observability windows associated with non-parametric morphological perturbations relative to close pairs. Furthermore, both methods are inherently biased toward specific physical properties of the interactions, such as stellar mass ratios, gas fractions, or orbital configurations. A more physically meaningful comparison would therefore involve merger rates, which incorporate these differing observability timescales into a unified framework.

\end{enumerate}

\section*{Data availability}

DEVILS: Data products used in this paper are taken from the internal DEVILS team data release and presented in \citet{Davies_2021} and \citet{Thorne_2021}. These catalogues were made public as part of the DEVILS first data release, as described in \citet{Davies_2025e}.

\section*{Acknowledgements}

We thank the reviewer for their comments which helped
improve this work. MFFF, LJMD, and ASGR, acknowledge support from the Australian Research Councils Future Fellowship schemes (FT200100055 and FT200100375). SB acknowledges support via an ARC Australian Laureate Fellowship (FL220100191).

DEVILS is an Australian project based around a spectroscopic campaign using the Anglo-Australian Telescope. DEVILS is part funded via Discovery Programs by the Australian Research Council and the participating institutions. The DEVILS website is \hyperlink{blue}{https://devilsurvey.org}. The DEVILS data are hosted and provided by AAO Data Central (\hyperlink{blue}{https://datacentral.org.au/}).



\bibliographystyle{mnras}
\bibliography{example} 




%
\appendix

\section{Individual visual classifications} \label{AppendixA}

In Section \ref{Visual}, we present different visual merger fractions based on the level of agreement between the four authors who participated in the visual classification of mergers. Here, we present the individual results of each classifier. This internal comparison is intended to demonstrate the degree of consistency among classifiers and to highlight the inherent subjectivity of the visual selection in identifying morphological disturbances at different redshifts.

\begin{figure*}
    \centering
	\includegraphics[scale=0.48]{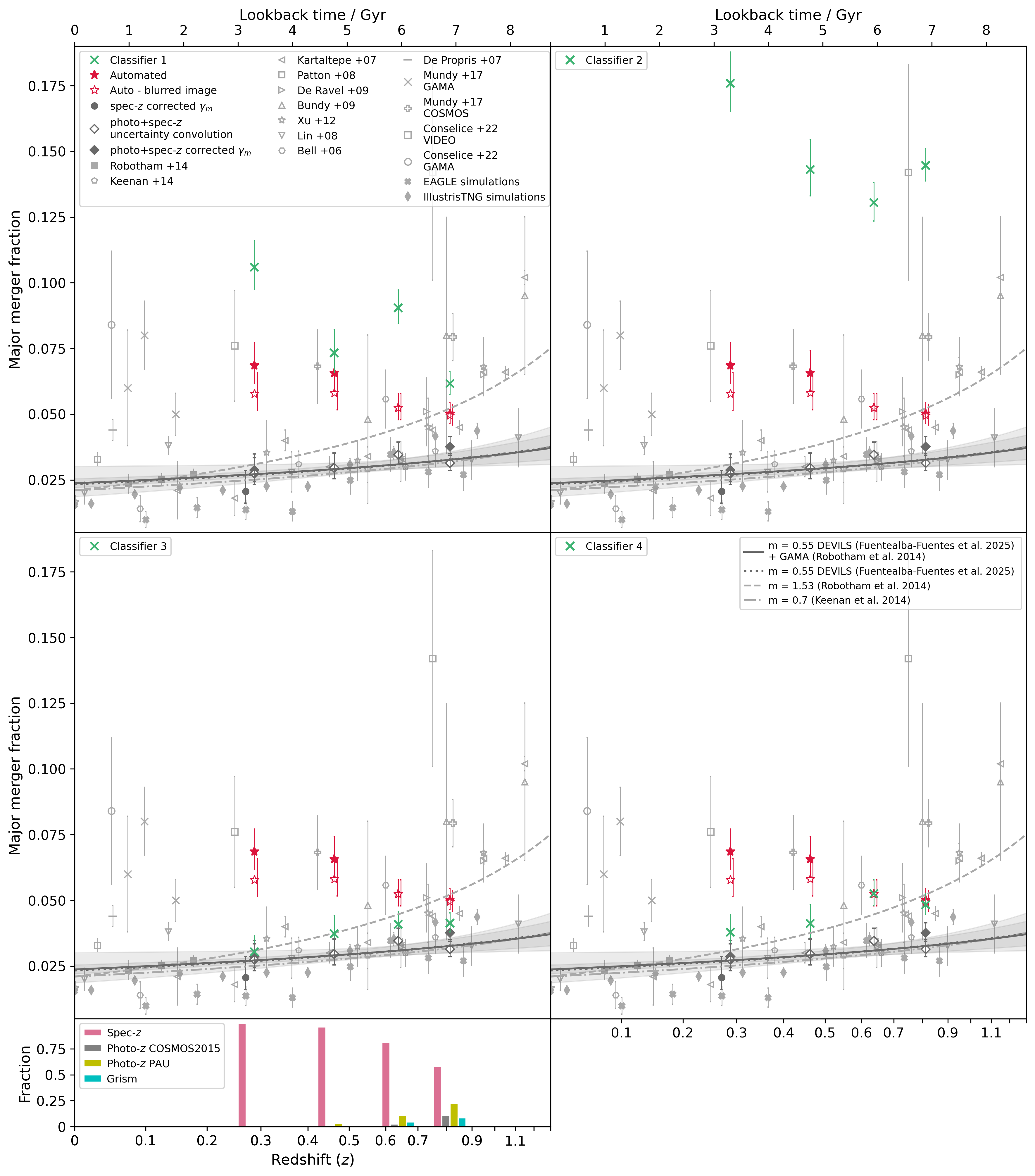}
    \caption{Galaxy merger fractions using the D10 (COSMOS) sample. Each panel shows the galaxy merger fractions estimated from the visual classifications of a single classifier (green). Symbols follow the same convention as in Figures \ref{fig:MergerFractionsVisual} and  \ref{fig:MergerFractionsAuto}: grey circles and diamonds show the close-pair merger fractions from \citet{Fuentealba-Fuentes_2025}, while filled and unfilled red stars indicate the automated classification of mergers using the original unsharped-masked image and the physical resolution matched $+$ unsharp-masked images, respectively. }
    \label{fig:Classifiers}
\end{figure*}

As shown in Figure \ref{fig:Classifiers}, we observe two different trends: one where the visual fractions are above the automated classifications and one where they fall below. The result that visual merger fractions are systematically higher than the automated estimates (derived from non-parametric statistics) is recovered once we account for the consensus across classifiers. A key finding of this work is that morphological merger fractions are consistently higher than the close-pair estimates from \citet{Fuentealba-Fuentes_2025}, a trend that is clearly observed in all four individual visual classifications.

\bsp	
\label{lastpage}
\end{document}